\documentclass[fleqn,10pt]{wlscirep}
\usepackage[utf8]{inputenc}
\usepackage[T1]{fontenc}
\usepackage{amsmath}
\title{Single-pass STEM-EMCD on a zone axis using a patterned aperture: progress in experimental and data treatment methods}

\author[1,*]{Thomas Thersleff}
\author[1, 2]{Linus Schönström}
\author[1]{Cheuk-Wai Tai}
\author[4]{Roman Adam}
\author[4]{Daniel E. Bürgler}
\author[4]{Claus M. Schneider}
\author[3]{Shunsuke Muto}
\author[2]{J\'{a}n Rusz}
\affil[1]{Stockholm University, Department of Materials and Environmental Chemistry, 10691 Stockholm, Sweden}
\affil[2]{Uppsala University, Department of Physics and Astronomy, Box 516, 75120 Uppsala, Sweden}
\affil[3]{Nagoya University, Institute of Materials and Systems for Sustainability, Nagoya 464-8603, Japan}
\affil[4]{Forschungszentrum Jülich GmbH, Peter Grünberg Institut, D-52425 Jülich, Germany}
\affil[*]{thomas.thersleff@mmk.su.se}

\begin{abstract}
Measuring magnetic moments in ferromagnetic materials with atomic column resolution is theoretically possible using the electron magnetic circular dichroism (EMCD) technique in a (scanning) transmission electron microscope ((S)TEM).  However, experimental and data processing hurdles currently hamper the realization of this goal.  Experimentally, the sample must be tilted to a zone-axis orientation, yielding a complex distribution of magnetic scattering intensity, and the same sample region must be scanned multiple times with sub-atomic spatial registration necessary at each pass.  Furthermore, the weak nature of the EMCD signal requires advanced data processing techniques to reliably detect and quantify the result.  In this manuscript, we detail our experimental and data processing progress towards achieving single-pass zone-axis EMCD using a patterned aperture.  First, we provide a comprehensive data acquisition and analysis strategy for this and other EMCD experiments that should scale down to atomic resolution experiments.  Second, we demonstrate that, at low spatial resolution, promising EMCD candidate signals can be extracted, and that these are sensitive to both crystallographic orientation and momentum transfer.
\end{abstract}
\begin{document}

\flushbottom
\maketitle
% * <john.hammersley@gmail.com> 2015-02-09T12:07:31.197Z:
%
%  Click the title above to edit the author information and abstract
%
\thispagestyle{empty}

%\noindent Please note: Abbreviations should be introduced at the first mention in the main text – no abbreviations lists. Suggested structure of main text (not enforced) is provided below.

\section*{Introduction}

Rapid developments in the field of nanotechnology call for experimental methods capable of providing information at sufficiently high spatial resolution. In the field of nano-magnetism, there are several well-established techniques such as x-ray magnetic circular dichroism \cite{chao_soft_2005} (XMCD), spin-polarized scanning tunneling microscopy \cite{wiesendanger_topographic_1992,heinze_real-space_2000}, magnetic exchange force microscopy \cite{kaiser_magnetic_2007} or electron holography \cite{midgley_electron_2009}. However, these methods lack either depth sensitivity or spatial resolution. An electron magnetic circular dichroism (EMCD) technique \cite{schattschneider_detection_2006}, a (scanning) transmission electron microscopy ((S)TEM) analogue of XMCD, in principle offers depth-sensitivity simultaneously with atomic spatial resolution.

EMCD has gone through a rapid development since its proposal in 2003 \cite{hebert_proposal_2003} and the first experimental demonstration in 2006 \cite{schattschneider_detection_2006}. It has been shown that the STEM geometry can be used\cite{schattschneider_detection_2008,salafranca_surfactant_2012} and that this can be exploited to both improve the signal to noise ratio (SNR) of an EMCD signal \cite{thersleff_quantitative_2015} as well as map magnetic moments in real space \cite{schattschneider_mapping_2012,thersleff_towards_2017}.  In the domain of high spatial resolution, EMCD has been detected using convergent beams of atomic size in a classical three-beam geometry \cite{thersleff_detection_2016}, utilizing phase ramps introduced by beam shift \cite{rusz_magnetic_2016}, in zone axis orientation \cite{song_detection_2016,spiegelberg_blind_2018}, and using atomic size beams distorted by four-fold astigmatism \cite{rusz_achieving_2014,idrobo_detecting_2016}. Recent experiments with a weakly convergent electron beam, using both geometric and chromatic aberration correction, led to a detection of EMCD from individual atomic planes \cite{wang_atomic_2018}. Despite some impressive achievements, EMCD technique is still under development, primarily due to the struggle with low magnetic signal strength and its sensitivity to dynamical diffraction effects and experimental artifacts. In an effort to overcome these difficulties, there is a need for both innovative experimental design and data analysis methods.

Recently, it was proposed to use patterned apertures for acquisition of EMCD signal \cite{negi_proposal_2019}. This approach is expected to bring several advantages. First, it enables acquisition of EMCD over the whole range of spatial resolutions, down to atomic scale. Second, it should offer a dose-efficient approach, because it utilizes a larger fraction of the inelastically scattered electrons than other acquisition geometries, thereby improving the signal-to-noise ratio (SNR) of the notoriously weak EMCD signal. Third, it offers a path to a single-pass STEM acquisition of the spectra, if the data from the whole CCD camera can be recorded at each scan point. Related experiments have demonstrated the feasibility of this approach \cite{schattschneider_magnetic_2008, ali_quantitative_2018}.

Here, we report our experimental progress towards single-pass STEM-EMCD on a cubic metallic iron sample using a patterned aperture as well as advances in data processing that are necessary to search for and extract potential EMCD signals.  This second part proves to be particularly challenging given the complex distribution of magnetic scattering on a zone-axis as well as the nature of geometric distortions present in 2D electron energy-loss spectroscopy (EELS) dispersion plane.  We begin with an outline of our hardware implementation including the design and installation of an 8-blade patterned aperture with a mirror symmetry plane oriented parallel to the spectrometer dispersion axis.  We then detail the acquisition scheme and describe the steps needed to reproduce the experimental method.  Subsequently, we summarize our analysis focusing on the impacts of 1) sample orientation and 2) momentum transfer in the non-dispersive axis (denoted $q_y$ in this manuscript).  We reveal that the most promising candidate EMCD signals are detected at high $q_y$ values extracted from regions of the sample that are oriented close to the Fe $\left [ 0 0 1 \right ]$ zone axis.  If a larger range of sample orientations are included in the analysis, we still observe a potential EMCD signal; however, its strength is diminished.  We also observe that the range of $q_y$ vectors plays a crucial role in this experiment, even observing an inversion of the signal sign on Fe $L_3$ that is not reciprocated on Fe $L_2$, which we interpret in terms of experimental shortcomings.  We conclude this manuscript with a discussion of these observations with accompanying theoretical considerations.  The data and code required to reproduce this analysis are provided and freely distributed\cite{thersleff_single-pass_2019}.  A version of the script formatted for publication detailing the analysis is provided in the supplementary information.

\section*{Results}

The experiment presented in this manuscript was designed to test two hypotheses that have been previously proposed.  The first hypothesis is that an 8-blade patterned aperture can detect an EMCD signal on the $\left [ 0 0 1 \right ]$ zone axis of bcc iron.  Given that simulations show that the EMCD signal is not very sensitive to the convergence angle and, thus, the ultimate spatial resolution of the probe\cite{negi_proposal_2019}, these experiments were performed with a probe width that was much larger than what would be required for probing atomic columns.  This allows the use of a more easily modified non C$_s$-corrected microscope configured with a large probe current of 2~nA to optimize the signal-to-noise ratio of the EELS spectra.

The second hypothesis is that, given the complexity of magnetic scattering on a zone-axis, the strength and, potentially, sign of a corresponding EMCD signal should depend strongly on the momentum transfer vectors that are allowed to pass to the spectrometer due to the patterned aperture.  This is expected to depend both on the rotation of the diffraction pattern with respect to the patterned aperture as well as the tilt of the crystal.  To test this, the TEM was operated in STEM mode and STEM-diffraction was performed to acquire convergent beam electron diffraction (CBED) patterns from the same region that was probed for EELS.  This allows for the crystallographic orientation to be locally extracted and correlated to EMCD signal strength and sign.  We achieve this by looking at the EMCD signal as a function of both $q_y$ and tilt away from the zone-axis.  Methodological details are provided in the methods section below.

\subsection*{Mirrored ventilator aperture}

\begin{figure}[ht]
	\centering
	\includegraphics[width=\linewidth]{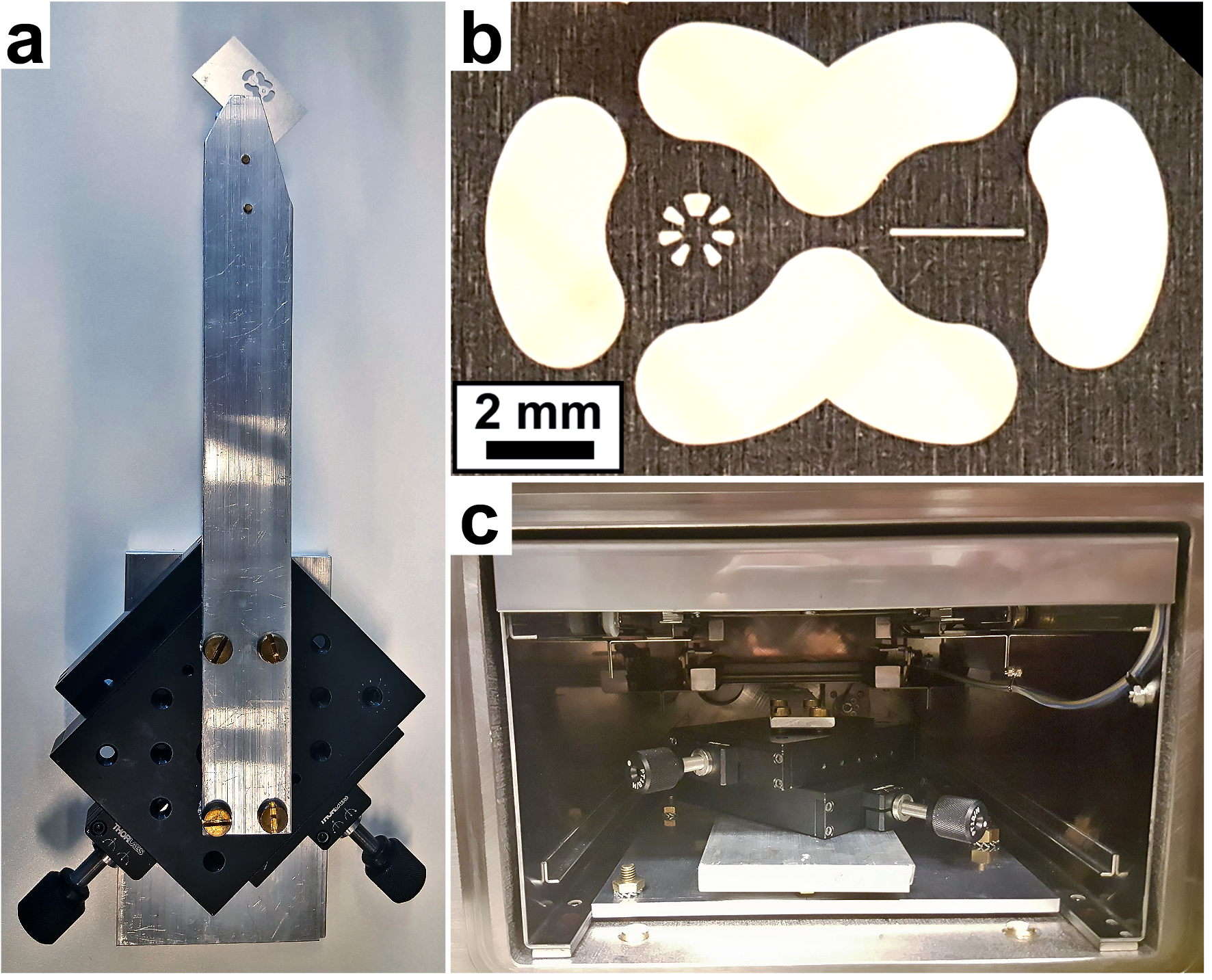}
	\caption{(a) Design of the aperture holding arm suspending the aperture into the beam path.  (b) The aperture plate.  This design includes one slit (right) and one 8-blade mirror-symmetry patterned aperture (left).  Switching between the two was accomplished by venting the chamber and adjusting the position.  The large annular holes permit the use of the lower HAADF detector.  (c) Placement of the aperture table into the TEM camera chamber.}
	\label{fig:ApertureDesign}
\end{figure}

% We need a paragraph explaining the aperture pattern (its 8 blades, not its mirror symmetry) and the ventilator moniker here!

\begin{figure}[ht]
	\centering
	\includegraphics[width=\linewidth]{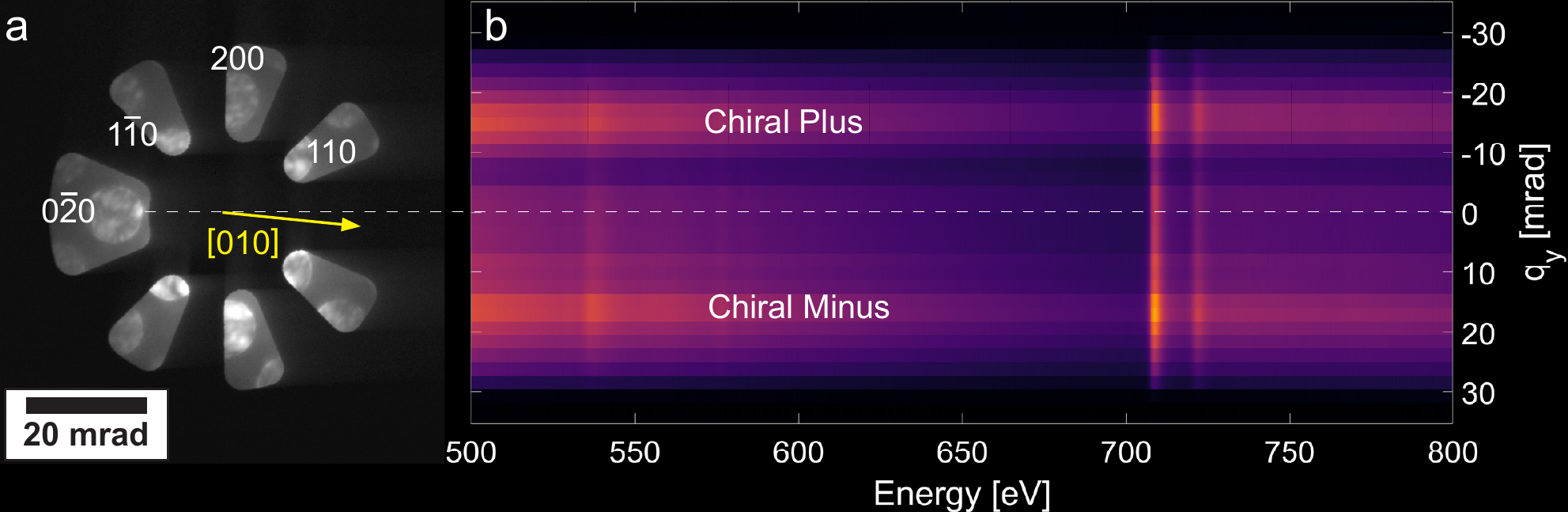}
	\caption{(a) Experimental diffraction pattern recorded on the GIF CCD showing the electron optical conditions.  The probe was centered on the the grain of interest (see Fig.~\ref{fig:4Dscans}).  (b) 2D EELS spectrum acquired using the electron optical conditions presented in (a).  The $q_y$ axis represents the momentum transfer that would be expected for a slit oriented normal to the spectrometer dispersion plane.  The oxygen edge visible at 532 eV comes from the Al$_2$O$_3$ capping layer and is primarily dominant in the background regions.}
	\label{fig:ApertureDispersion}
\end{figure}

The pattern used for this aperture is presented on the left side of Fig.~\ref{fig:ApertureDesign}b.  It is an 8-blade mirror-symmetry design optimized for single-pass EMCD acquisition on the $\left [0 0 1 \right ]$ zone axis of a cubic material, as proposed by Negi et al. \cite{negi_proposal_2019}.  In this figure, the energy dispersion axis is oriented vertically.  The right side of Fig.~\ref{fig:ApertureDesign}b shows a slit aperture that was not used in this experiment.  The large holes surrounding the pattern allow for electrons to pass through the aperture plate to the high-angle annular dark field (HAADF) detector located at the spectrometer entrance.  This aperture was suspended in the beam path by an electrically grounded arm mounted on a movable table, as shown in Fig.~\ref{fig:ApertureDesign}a.  The table allowed for fine adjustments of the aperture position prior to evacuation of the camera chamber.  The entire system was made vacuum compatible and placed into the negative chamber of a JEOL 2100F TEM, as shown in Fig.~\ref{fig:ApertureDesign}c.  Adjustments were made by opening the chamber door, requiring venting and subsequent evacuation of the camera chamber.  The aperture position on the Ultrascan camera in the EELS spectrometer is presented in Fig.~\ref{fig:ApertureDispersion}a.  We note that, unlike the rotation of the diffraction pattern, the alignment between the patterned aperture and the EELS spectrometer is fixed and, thus, critical to be correctly set prior to the experiment.

Following adjustment of the aperture position, a suitable iron grain was sought out and brought into the field of view following the methodology presented in the methods section.  The goniometer was tilted to align the grain along the $\left [ 0 0 1 \right ]$ zone axis and the diffraction pattern was rotated using the projector system of the TEM to align the $\left [ 0 1 0 \right ]$ axis with the spectrometer dispersion axis.  Figure \ref{fig:ApertureDispersion}a shows the CBED pattern from this orientation as well as its associated rotation to the patterned aperture, allowing for exact calculation of the collection angles.  The probe position for this pattern was close to the center of the grain.

The spectrometer was operated in frame acquisition mode and a series of 2D EELS spectra were acquired using the experimental conditions detailed in the methods section.  A summation over all of the 2D EELS images following alignment along the energy axis is presented in Fig.~\ref{fig:ApertureDispersion}b.  The iron edges are visible as vertical streaks at 709 and 723~eV.  This $q-E$ diagram is aligned vertically with the aperture placement on the CBED pattern in Fig.~\ref{fig:ApertureDispersion}a, illustrating the experimental concept.  

%Of particular importance in Fig.~\ref{fig:ApertureDispersion}a is how the non-dispersion dimension ($q_y$) interacts with the patterned aperture.  To a first approximation, one can consider that the passage of the CBED pattern through the magnetic prism of the spectrometer "collapses" the $q_x$ dimension, while the $q_y$ dimension remains largely intact.  This creates a superposition of scattering momenta along the $q_x$ dimension onto the $q_y$ position in the spectrometer.  Consequently, we must consider the $q_y$ values close to 0 to consist of a superposition of a wide range of $q_x$ vectors, whereas the higher $q_y$ values will be more selective in the probed $q_x$ range.

Of note here is how the shape and orientation of the mirrored ventilator aperture determines the regions of scattering angles collected by the detector. The $q_x$ dimension is integrated out, leaving only $q_y$ and $E$ dependences in the acquired dataset. Considering the shape of the mirrored ventilator aperture, the $q_y$ values close to 0 thus consist of a sum over a wide range of $q_x$, whereas the higher $q_y$ values will be more selective in the probed $q_x$ range.

\subsection*{4D-STEM experiments}

\begin{figure}[ht]
	\centering
	\includegraphics[width=\linewidth]{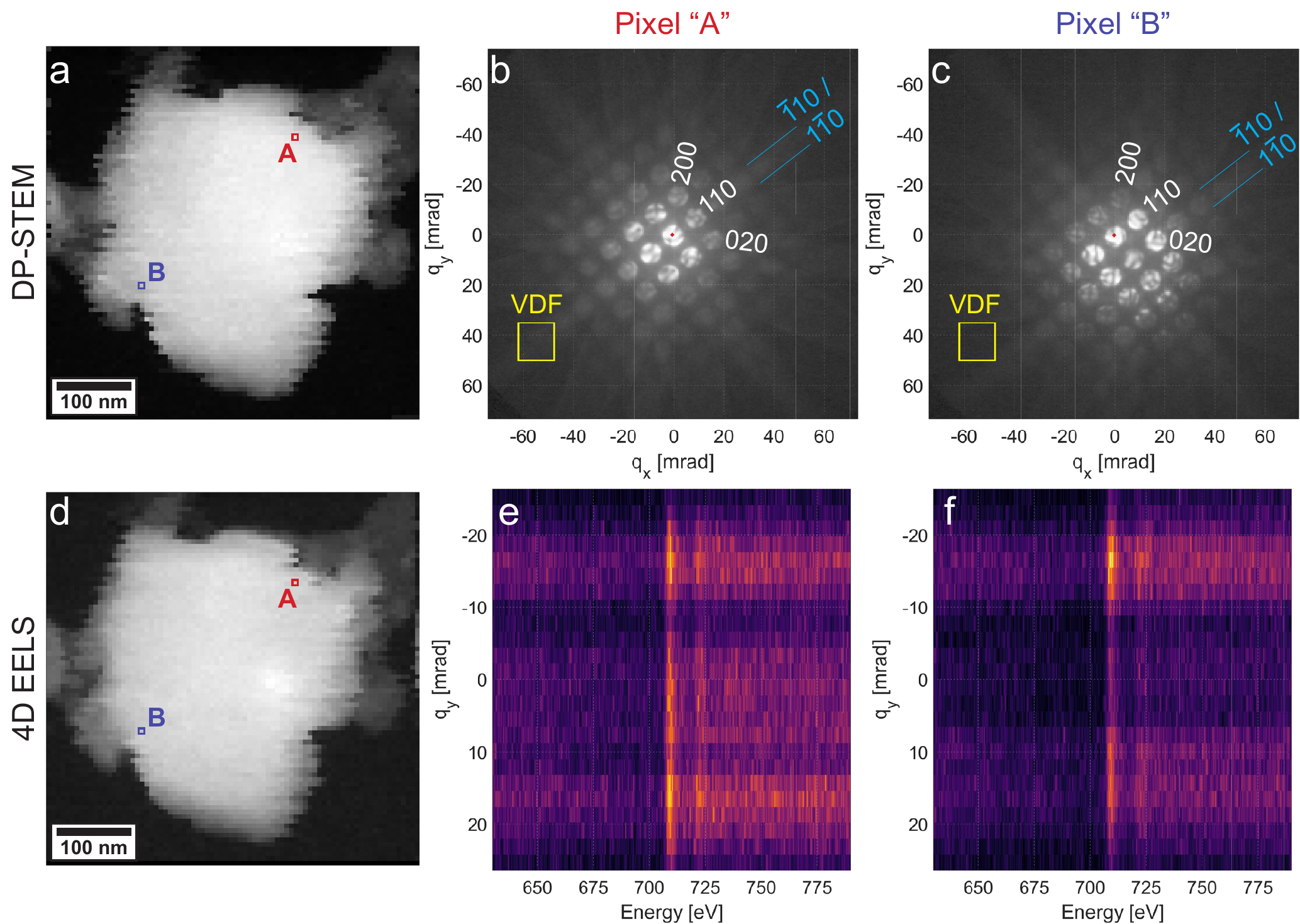}
	\caption{(a) Virtual HAADF of the 4D STEM datacube over the grain of interest.  The CBED patterns acquired at pixel position "A" and "B" are shown in (b) and (c), respectively.  A gamma curve of 0.3 is applied to the grayscale mapping in these two images to enhance the visibility of the otherwise very weak Kikuchi lines, which are indexed and highlighted in cyan.  The center of the $000$ reflection is denoted with a red dot. The virtual aperture used to generate the VDF micrograph in figure \ref{fig:ROIselection}c is also labeled and presented in yellow.  (d) Summation over $\Delta E$ and $q_y$ for the 4D EELS datacube.  The 2D EELS spectra from pixel positions "A" and "B" are presented in (e) and (f), respectively.}
	\label{fig:4Dscans}
\end{figure}

The 4D datacubes from both the CBED patterns and the 2D EELS images are summarized in Fig.~\ref{fig:4Dscans}.  The STEM-DP 4D datacube reveals that the orientation of the probed iron grain varies as a function of the probe position.  This variation may have been caused by a slight beam tilt induced while scanning such a large area, as this TEM is not equipped with descan coils.  This tilt effect is exemplified in Fig.~\ref{fig:4Dscans}a-c.  Two representative pixel positions ("A" and "B") were chosen from this dataset, as displayed in Fig.~\ref{fig:4Dscans}a.  The corresponding CBED patterns are displayed in Fig.~\ref{fig:4Dscans}b and c, respectively.  This figure summarizes the more interactive experience of visualizing all of the collected CBED patterns as a function of probe position, which was performed using the PyXem plugin for Hyperspy \cite{francisco_de_la_pena_hyperspy/hyperspy:_2018,johnstone_pyxem/pyxem:_2019}.  The qualitative interpretation of this interactive experience was that the "upper-right corner" of the grain was oriented closer to the zone axis.  Moreover, the grain appears to primarily rotate about the $\left [ 1 1 0 \right ]$ axis.  This rotation can be visualized by placing a virtual aperture away from the Bragg disks but centered on the $\overline{1} 1 0 / 1 \overline{1} 0$ Kikuchi line pair for the zone-axis orientation, as shown with the yellow box in Fig.~\ref{fig:4Dscans}b,c.  This is illustrated in Fig.~\ref{fig:ROIselection}c and is discussed more quantitatively in the next section.

%The subsequent contrast generated in the virtual dark field (VDF) image thus gives a rough approximation of the degree of mistilt (see Fig.~\ref{fig:ROIselection}c).

It should be noted that, since these datacubes were acquired using custom written hook-up scripts with the spectrum imaging subsystem in Digital Micrograph \cite{schonstrom_time_2018}, the synchronization between the probe position and the data acquisition is not perfect.  A spatial distribution of the number of frames acquired per pixel is provided in the supplementary information.  While most pixels contain one dataset, many contain zero or two.  Nearest neighbor interpolation was used to fill the $80 \times 80$ pixel field of view.  

\subsection*{Orientation mapping and selection of region of interest}

\begin{figure}[ht]
	\centering
	\includegraphics[width=12cm]{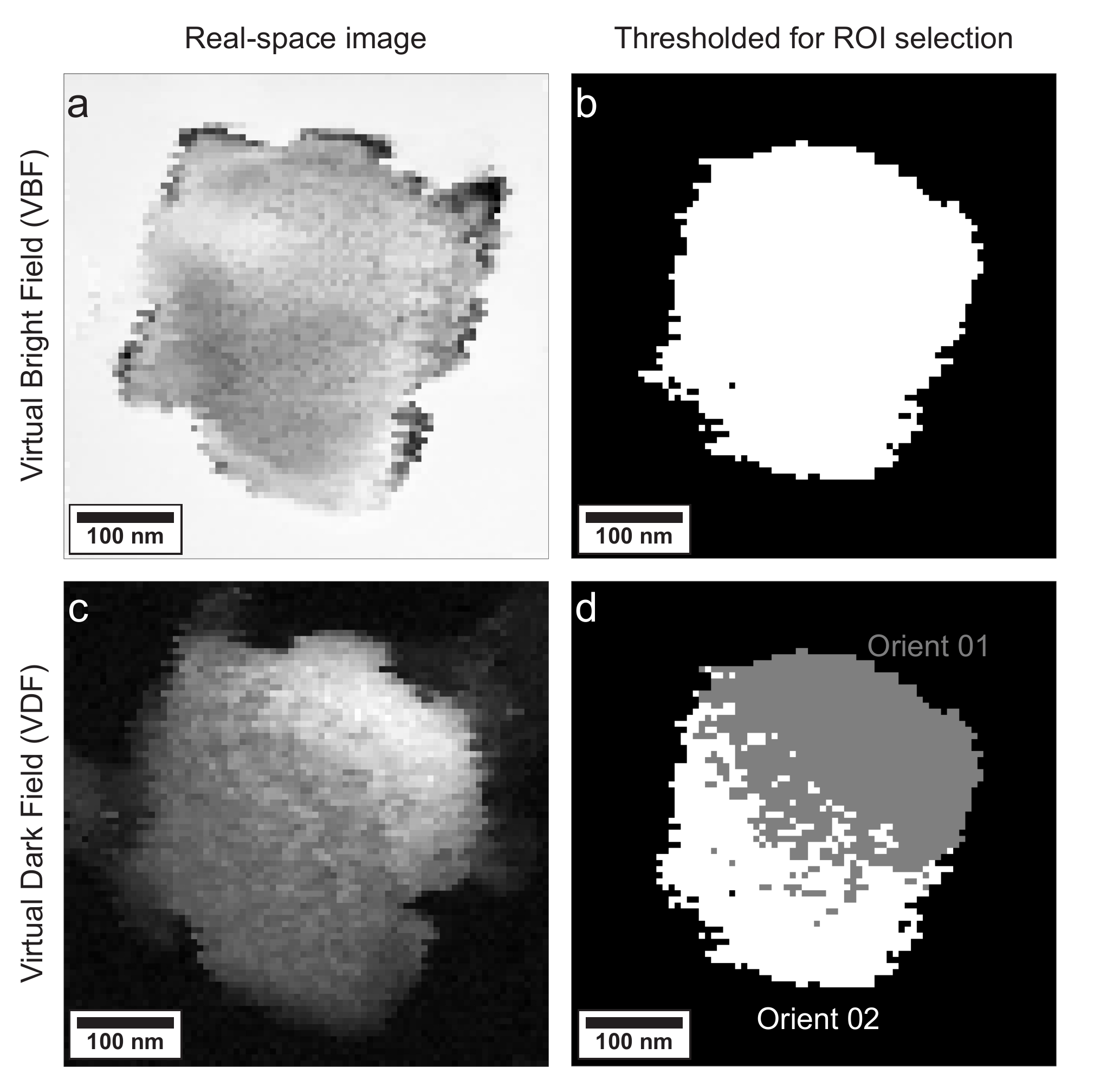}
	\caption{(a) Real-space micrograph created by using a virtual bright field aperture.  (b) ROI selection using a threshold for the full grain mask. (c) Real-space micrograph created by using the virtual dark field aperture presented in figure \ref{fig:4Dscans}.  (d) The two orientation masks, "Orient 01" and "Orient 02" were generated by thresholding this image, and both are composited here.  The exact thresholding parameters are provided in the supplementary information.}
	\label{fig:ROIselection}
\end{figure}

To test the hypothesis that EMCD signal strength should vary with zone-axis mistilt, virtual bright field (VBF) and virtual dark field (VDF) images were generated using the STEM-DP datacube.  The VBF image was formed by integrating over a box centered in the transmitted disk, and is presented in Fig.~\ref{fig:ROIselection}a.  This effectively measures the loss of intensity to additional Bragg disks in the CBED patterns.  Since the central grain was oriented along a zone-axis, significant intensity is lost and the grain itself appears dark.  This image was thresholded to reveal a selection mask that restricts the region of interest to the zone-axis oriented grain, as presented in Fig.~\ref{fig:ROIselection}b and the supplementary information.

Within this grain, a rotation about $\left [ 1 1 0 \right ]$ is observed, as discussed above.  An estimate of the degree of this rotation (corresponding to a mistilt) can be obtained by generating a VDF image using the virtual aperture shown in Figs.~\ref{fig:4Dscans}b and c.  The resulting VDF is presented in Fig.~\ref{fig:ROIselection}c.  A significant increase in the intensity values of the pixels in this image thus represents, to a first approximation, an orientation that is closer to the $\left [ 0 0 1 \right ]$ zone-axis geometry.  

This VDF was qualitatively thresholded using an empirically-determined value to yield a spatial mask segmenting the grain scattering geometry close to a zone-axis and off of a zone-axis (details are provided in the supplementary information).  The two regions of interest (ROI) are presented in Fig.~\ref{fig:ROIselection}d.  We note that this segmentation is not intended to be quantitative: it merely serves as a way to test the hypothesis of whether there is a dependence of the EMCD signal strength on the sample orientation.  The mask labeled "Orient 01" refers to the pixels where the CBED patterns show a closer to the zone-axis orientation, while "Orient 02" refers to pixels with a stronger tilt away from the zone-axis geometry.  These orientation masks, as well as the "grain" mask from figure \ref{fig:ROIselection}b, are used in the EMCD signal extraction below.

\subsection*{Candidate EMCD spectra}

%Extraction of EMCD signals from this dataset requires significant data processing.  A detailed description of the processing steps taken to extract the following candidate EMCD signals is provided in the methods section below.  The Matlab code and raw data are provided as well and can be downloaded from Zenodo \cite{thersleff_experimental_nodate}.

In this section, the two hypotheses described above are tested.  The null hypothesis is defined as "no EMCD signal is detected."  As a detection criterion, we use the methodology outlined in Thersleff et al. \cite{thersleff_detection_2016}, where the confidence in a positive EMCD signal detection is expressed as the SNR of both the Fe $L_3$ and $L_2$ edges.  Asserting the Rose criterion, a SNR of 5 (i.e. 7~dB) or more on \textit{both} the Fe $L_3$ and $L_2$ edges is necessary to confidently reject the null hypothesis (that no EMCD signal is present).  EELS difference spectra with positive SNR, but lower than the Rose criterion, will be described here as "candidate" EMCD spectra, signifying the degraded confidence.  Since the $L_2$ edge is more difficult to resolve, this will be the primary focus in the following discussion.  

%\begin{figure}[ht]
%	\centering
%	\includegraphics[width=10cm]{Figures/EMCD_SignalCandidates.png}
%	\caption{EMCD signal candidates taken from $q-E$ of high $q_y$ and (a) orientation mask 1 and (b) orientation mask 2.  The EMCD signal in $L_2$ in both is on the order of 1\%.  EMCD signal candidates from the grain mask and $q-E$ regions of (c) high $q_y$ and (d) low $q_y$.  An potential EMCD signal is visible in (c) but is clearly absent in (d).}
%	\label{fig:SignalCandidates}
%\end{figure}

\subsubsection*{Influence of zone-axis mistilt}

\begin{figure}[ht]
	\centering
	\includegraphics[width=15cm]{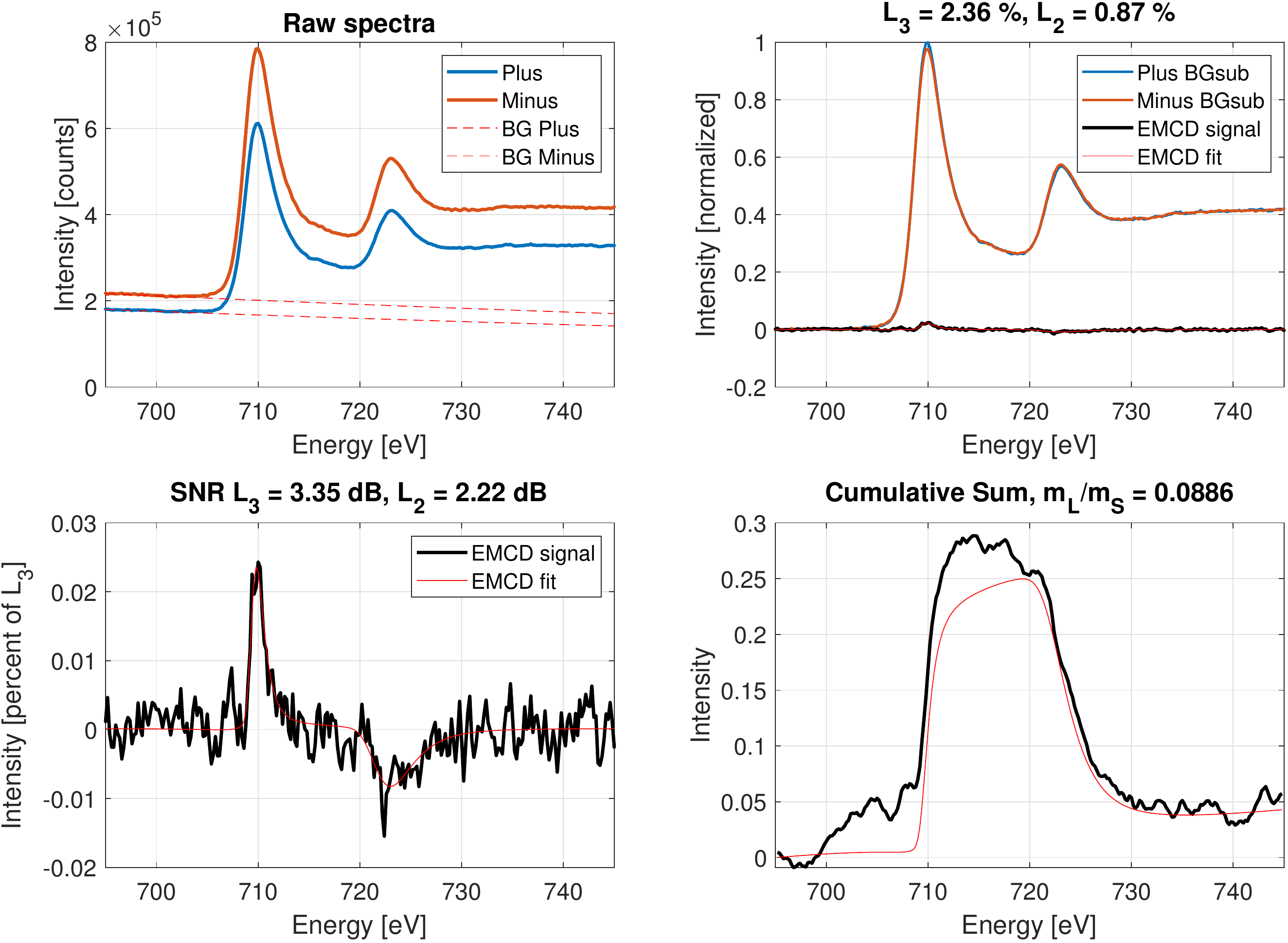}
	\caption{Candidate EMCD signal taken from the outer-most $q_y$ range and the spatial mask restricted to the region oriented most close to the Fe [1~0~0] zone-axis.}
	\label{fig:EMCD_Orient01_Sum1-4}
\end{figure}

We begin by examining the influence the zone-axis mistilt has on the presence and sign of an EMCD signal.  For this, we first use the spectra collected from the spatial region closest to the zone-axis orientation, denoted by the orientation mask 01 (see Fig.~\ref{fig:ROIselection}d).  We also restrict ourselves to the outer $q_y$ range (17.6 -- 24.2~mrad), as will be justified below.  The candidate EMCD signal extracted from this region is presented in Fig.~\ref{fig:EMCD_Orient01_Sum1-4}.  Despite a relatively weak signal strength of less than 1\% on Fe $L_2$, the SNR is 2.2~dB using these extraction settings.  While this does not meet the Rose criterion, this gives a decent level of confidence that an EMCD signal can be detected in the data using this combination of extraction method, spatial sampling, and $q_y$ range.  %We note that an anomaly in the energy range 700 - 705~eV is observed here.  This is discussed in detail in the supplementary information and can be attributed to peak broadening caused by residual spectrometer aberrations. 

\begin{figure}[ht]
	\centering
	\includegraphics[width=15cm]{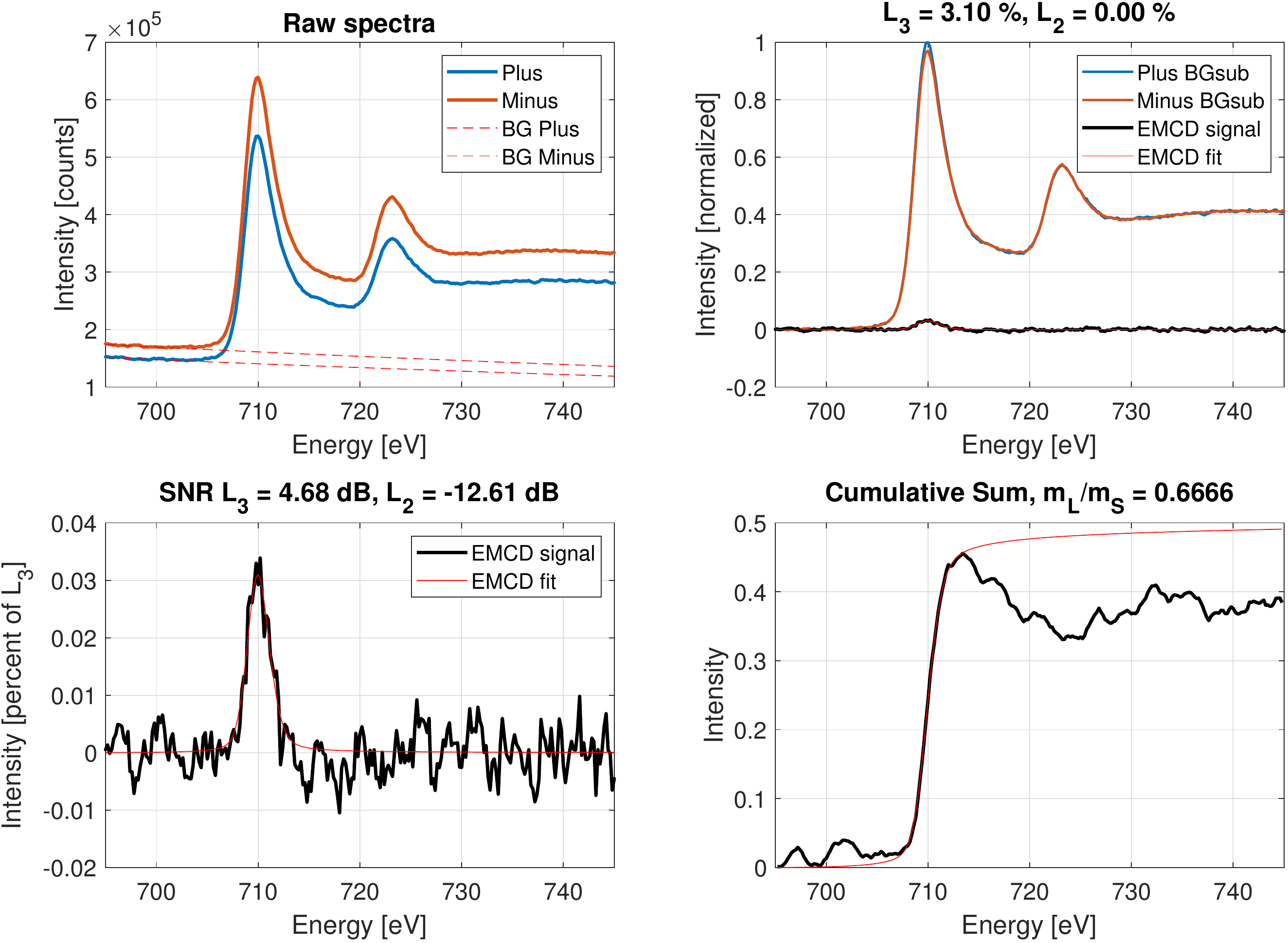}
	\caption{Candidate EMCD signal taken from the outer-most $q_y$ range and the mask restricted to the region oriented furthest from the Fe [1~0~0] zone axis.}
	\label{fig:EMCD_Orient02_Sum1-4}
\end{figure}

We now compare this signal to the one produced using the same $q_y$ integration range but collected from the spatial region exhibiting a larger mistilt from the zone-axis orientation, denoted by the orientation mask 02 (see Fig.~\ref{fig:ROIselection}d).  The EELS difference signal in this region, presented in Fig.~\ref{fig:EMCD_Orient02_Sum1-4}, shows a pronounced feature on Fe $L_3$ but no signal is visible on Fe $L_2$.  

\subsubsection*{Influence of $q_y$}

\begin{figure}[ht]
	\centering
	\includegraphics[width=15cm]{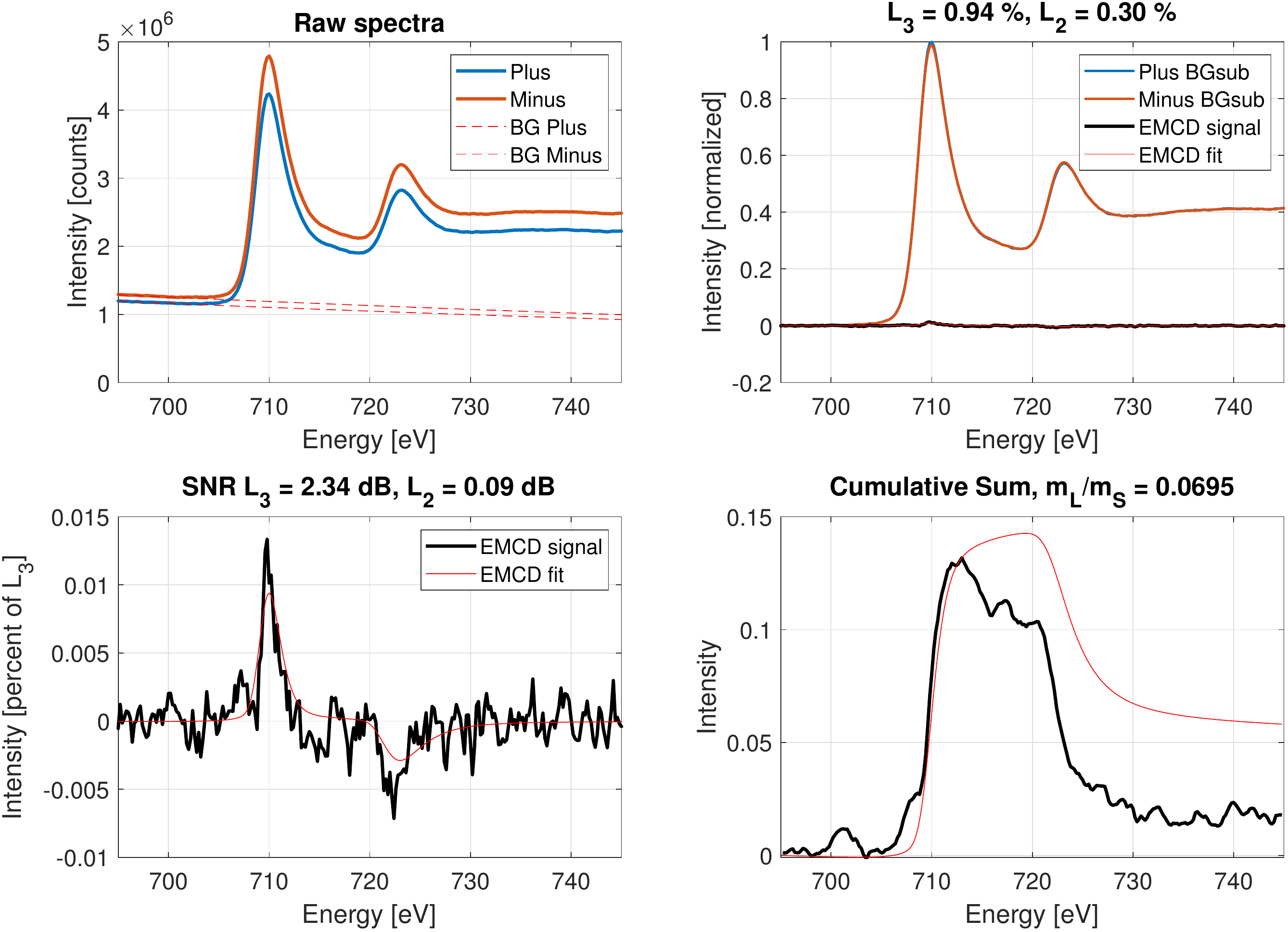}
	\caption{Candidate EMCD signal taken from the full $q-E$ range and the mask covering the entire zone-axis oriented grain.}
	\label{fig:EMCD_OrientGrain_SumAll}
\end{figure}

We now shift the focus from the zone-axis mistilt to the influence of $q_y$.  We begin by looking at all mistilt angles using the "grain" mask presented in Fig.\ref{fig:ROIselection}b and integrating over all $q_y$ values.  This describes the scenario most consistent with the theory and simulations from in Negi et al. \cite{negi_proposal_2019}.  The result of the EMCD signal extraction is presented in Fig.~\ref{fig:EMCD_OrientGrain_SumAll}.  In this case, a weak candidate EMCD signal appears to be visible with a signal strength of 2.3~dB on Fe $L_3$.  A weak but noticable signal is visible on Fe $L_2$ here as well, particularly when the cumulative sum of the EMCD signal is computed.  Given the results of zone-axis mistilt (Figs.~\ref{fig:EMCD_Orient01_Sum1-4} and \ref{fig:EMCD_Orient02_Sum1-4}), the comparatively weak Fe $L_2$ signal strength may be partially explained by the inclusion of various off-axis samples.

\begin{figure}[ht]
	\centering
	\includegraphics[width=15cm]{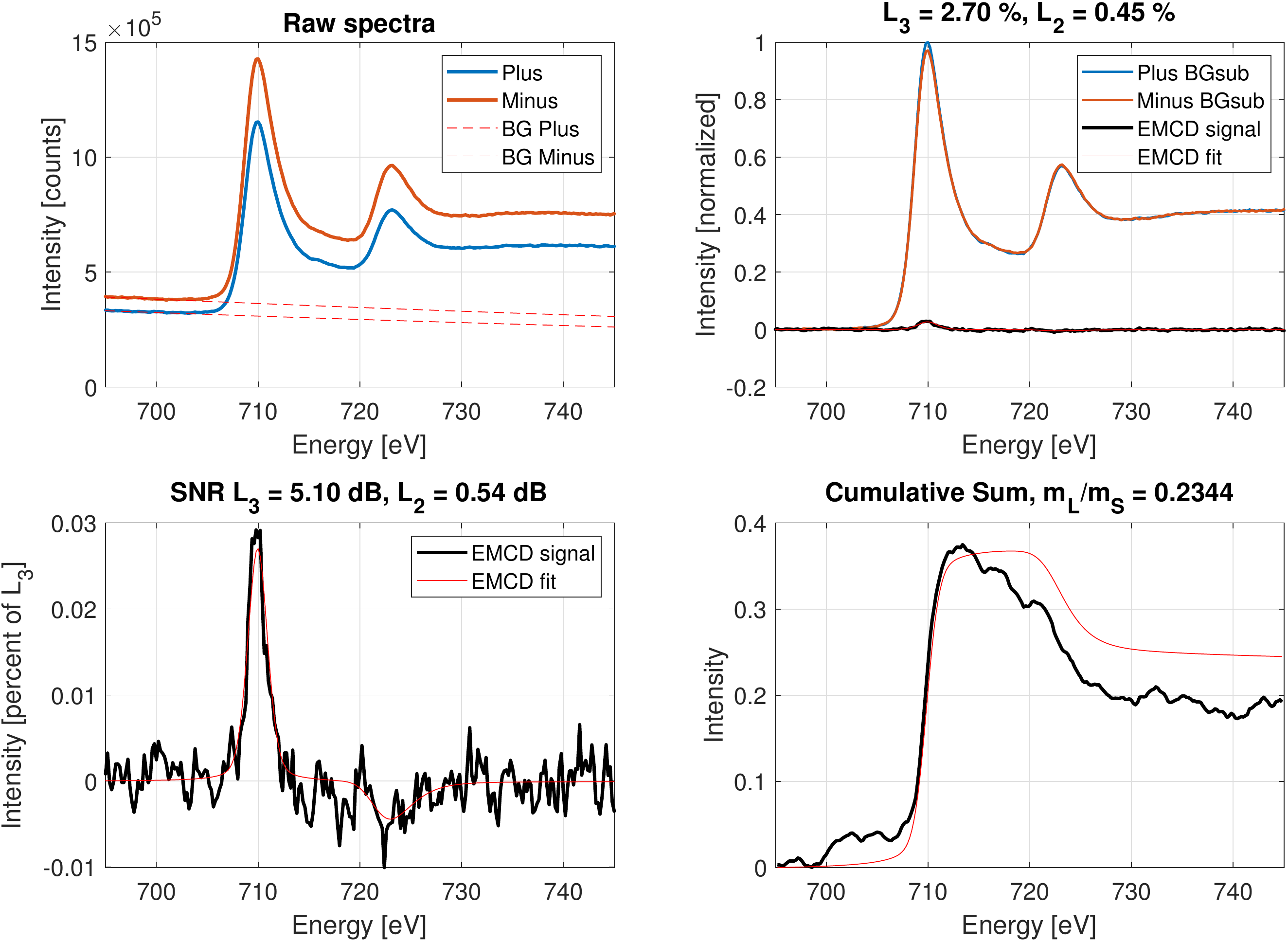}
	\caption{Candidate EMCD signal taken from the outer-most $q_y$ range and the mask covering the entire zone-axis oriented grain.}
	\label{fig:EMCD_OrientGrain_Sum1-4}
\end{figure}

We can explore this further by restricting the $q_y$ range to 17.6 -- 24.2~mrad, as above.  Using the same ROI mask from Fig.\ref{fig:ROIselection}b, we observe the results presented in Fig.~\ref{fig:EMCD_OrientGrain_Sum1-4}.  The strength of Fe $L_2$ has now increased to 0.5~dB, indicating a slightly improved confidence.  This suggests that the $q_y$ range does influence the extracted EMCD signal, with higher $q_y$ values (representing an integration over fewer $q_x$ values) yielding a more convincing EMCD signal.

%Our proposed solution is to parameterize a sharpening function and add this to the optimization routine.  Failure to perform this step leads to a difference signal that is clearly dominated by peak broadening effects and cannot be used to test the null hypothesis.  This step has not yet been debated in the EELS literature, but we argue in favor of its use here.  We base this argument largely on the observation that the integral of the second derivative of the spectrum to be sharpened exhibits a zero-crossing at around 718~eV, indicating the area under the $L_{2,3}$ peaks does not change and, hence, does not bias the EMCD signal extraction.  More details and code are provided in the methods section and supplementary information.  

%We now investigate the effect of restricting the spectra to the spatial region oriented furthest away from the zone-axis orientation, denoted by the orientation mask 02 (see Fig.~\ref{fig:ROIselection}d).  

\begin{figure}[ht]
	\centering
	\includegraphics[width=15cm]{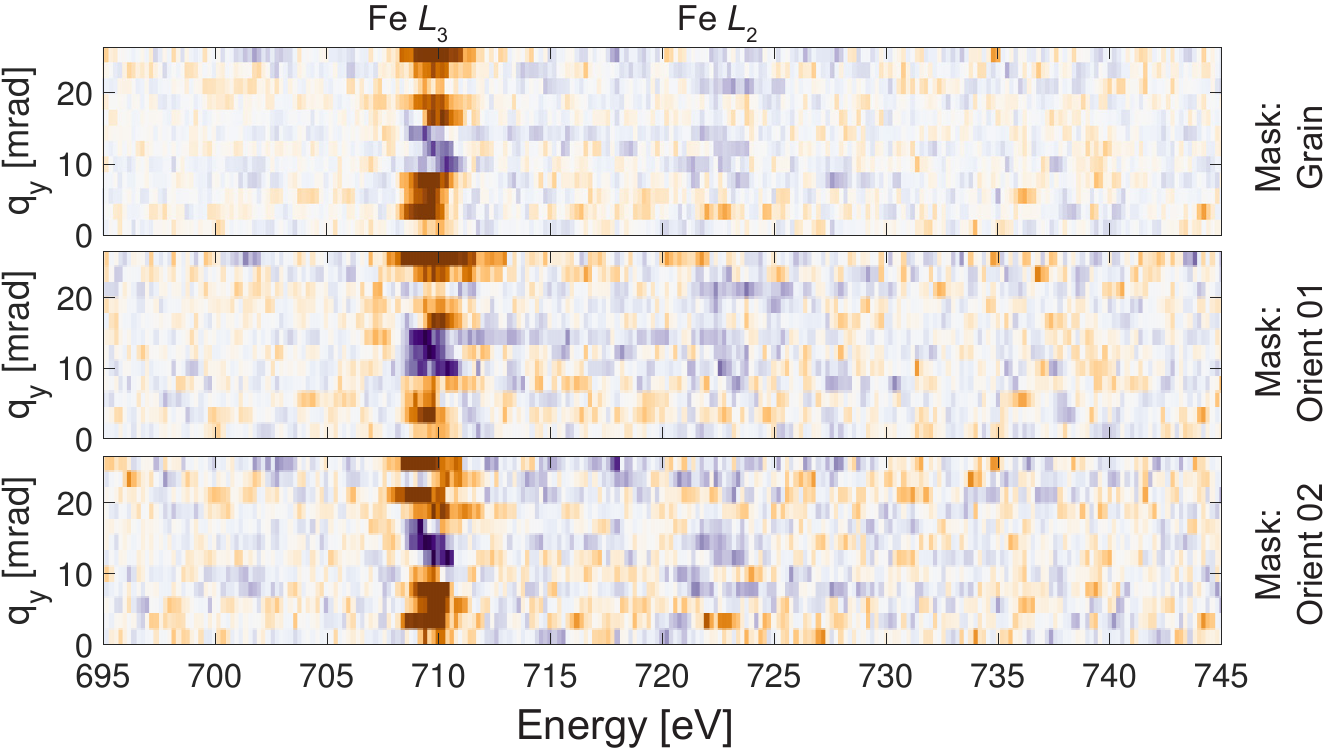}
	\caption{$q_y$ EMCD maps for the three different ROI masks.}
	\label{fig:qy_maps}
\end{figure}

We can study the effect of $q_y$ further by plotting the EMCD signal as a function of $q_y$ for each of the three ROIs masks.  This is presented in Fig.~\ref{fig:qy_maps}.  While the data are much noisier than with the previous figures (since no $q_y$ integration was performed), a weak signal on Fe $L_2$ is still visible for some $q_y$ values.  Most significant in this visualization, however, is the observation that the EELS difference signal on $L_3$ clearly flips sign for middle $q_y$ values.  The sign on $L_2$ does not appear to change.  Inspection of Fig.~\ref{fig:ApertureDispersion}a reveals that the $q_y$ range where flipping occurs has strong non-magnetic contributions from the $(110)$ Bragg reflections, which are likely to enhance the noise and thus reduce the magnetic SNR.

\section*{Discussion}

%The patterned aperture design with a mirror symmetry axis oriented parallel to the spectrometer dispersion plane reveals some promising candidate EMCD signals and, thus, has potential to become an effective technique for measuring EMCD in STEM mode in a single pass on the zone axis of a magnetic crystal.  As theory suggests that an EMCD signal remains even at high convergence angles, the potential for very high spatial resolution measurements is not precluded, provided that aberration corrected probes can be employed.  However, the use of broader probes should not be overlooked, as this can provide a powerful means for scanning large sample areas quickly and reliably while keeping certain grain boundaries edge-on.

Overall, the results of this experiment lead us to an optimistic assessment of the single-pass STEM-EMCD patterned aperture approach.  While none of the observed signals meet the Rose criterion, some signals of statistical significance are nevertheless present.  We also feel that this study has been particularly instructive at highlighting a number of limitations and outlining the challenges for this experiment, spurring the development of considerable data processing methods.  Here, we discuss the observed trends in the data as well as the novel data processing techniques that we have developed that were necessary for us to draw our conclusions.

The first trend we discuss is the influence of zone-axis mistilt.  We do observe a reduction in the strength of Fe $L_2$ for sample regions that are tilted further away from the zone-axis orientation.  Importantly, this does not seem to come at the expense of a signal on Fe $L_3$.  The loss of the Fe $L_2$ edge is something that has been observed in classical EMCD under due to partial mixing of the magnetic and non-magnetic signals resulting from an asymmetrical orientation along the systematic row \cite{rusz_asymmetry_2010,song_effect_2014}.  We may be similarly observing this mixing effect here, but in the zone-axis geometry.  The clear lesson to be learned from this is that the EMCD signal is likely to be highly sensitive to sample orientation, requiring minimal beam tilt during scanning operations (potentially limiting the field of view) and high precision with the initial tilting settings.

The second observed trend is the influence of $q_y$, with the sign inversion on the Fe $L_3$ edge for $q_y$ values approximately between 10 to 15~mrad being particularly intriguing (see Fig.~\ref{fig:qy_maps}). While a sign inversion is not observed for simulations with a perfectly oriented mirrored ventilator aperture \cite{negi_proposal_2019}, such sign inversions would be theoretically possible for a small rotation of the $\left [ 0 1 0 \right ]$ direction in the diffraction pattern from the mirror symmetry axis (which is physically aligned to the spectrometer energy dispersion axis). We observe such a rotation in this experiment, as indicated in Fig.~\ref{fig:ApertureDispersion} by the yellow arrow. In addition, small mis-alignments in the positioning of the aperture naturally lead to an asymmetry of the upper and lower diffraction half-plane (with respect to the mirror axis of the aperture), which, like the mistilt in the grain orientation, can also lead to further mixing of the magnetic and non-magnetic signals \cite{song_effect_2014,rusz_asymmetry_2010}. It is well possible that failure to achieve a near-perfect alignment of the crystal symmetry directions with the spectrometer is among the strongest factors influencing the EMCD extraction. Hardware limitations of the TEM instrument used in this work make a better alignment extremely tricky. However, more stable microscope columns equipped with more projector lenses and deflectors should greatly simplify this alignment step.  

In addition to these experimental considerations, a serious challenge originates from the residual aberrations in the EELS spectrometer.  Inspection of the raw data reveal that a number of spectral aberrations such as spectral blurring, energy offset, and even dispersion variations persist in the $q_y$ dimension of the $q-E$ diagram.  Given the complex distribution of magnetic scattering in the zone-axis geometry, such artifacts provide a major hurdle to the reliable extraction of an EMCD signal.  While reasonable measures were taken to ensure that the off-axis dispersion was optimal, the non-standard approach taken in this study requires such adjustments to be made without the use of computerized automation.  Consequently, spectral focus and astigmation were performed by hand.  Although deemed to be sufficient when operating the TEM, the extremely stringent demands of EMCD signal extraction amplify even the smallest of alignment errors.  

The approach we have adopted to meet these challenges at this stage is the development of more robust data processing techniques.  Some of the steps necessary to extract candidate EMCD signals in this paper have not yet been debated in the EELS literature in detail.  For this reason, we devote space here to outlining and justifying these steps, with the hope to stimulate a wider discussion on the topic. Simultaneously, we express hopes that future improved spectrometer hardware and automatic alignment procedures will remove the need of most (if not all) of these post-processing steps, as this would improve the quantitative performance over the whole range of EELS-based studies.

The first non-standard signal processing step that appears to be necessary for extracting an EMCD signal from these data is the addition of an energy dependence to the post-edge normalization.  This seems to be a critical step: failure to use this approach results in a linear trend in the post-edge regions.  Consequently, when we use a simple integral normalization, we find that the position of the post-edge window becomes overcritical, with the potential for human bias.  The energy dependence is kept as simple as possible: a simple linear trend is regressed to the energy-dependent ratio between the chiral plus and minus spectra in the post-edge region.  This approach has been taken previously by Schneider et al. \cite{schneider_magnetic_2016} (inspired from discussions within the XMCD community).  We believe that this normalization routine is not only less prone to such bias, but can be justified on physical grounds as spectrometer aberrations mentioned above.  In particular, we note that there is a slight curvature in the dispersion of the EELS spectra that bends them down along the increasing $E$ direction.  This may result in a linear post-edge trend, which we do, in fact, observe.  While we are not yet certain about how this correction affects the quantification of magnetic moments, we observe that it is crucial in extraction of the EMCD candidate spectra presented in the Results section. For a routine deployment of this processing step, more detailed analysis is necessary than what can be provided in this work. Alternatively, hardware or software correction procedures to minimize the bending distortion need to be developed.
%observe that it improves confidence of our qualitative detection criterion.  Despite this justification, we would like to emphasize that, without this step, the same conclusions in this report cannot be drawn.  We thus encourage the utility of this procedure to be debated in the wider community.

The second non-standard step involves the profile matching of the Fe $L_{2,3}$ edge shapes for both the chiral spectra via a peak broadening / sharpening function.  This removes the influence of spectral broadening that we observe as a function of the $q_y$ dimension.  This appears to be necessary due to residual geometrical aberrations in the spectrometer causing difficulties in achieving perfect focus along the entire $q_y$ dimension. Since quantitative EMCD utilizes spectral integrals, profile matching will not modify the quantification of an EMCD signal as long as the area under the peaks is invariant.  We achieve this by applying the standard signal processing technique of subtracting a multiple of the smoothed second derivative of the spectrum from the spectrum itself.  The peak profiles are matched using the optimization routine detailed in the Methods section below, and the step is parameterized.

%a sufficient goal of this step is to \emph{match} the broadening of the two chiral spectra, rather than to perform their \emph{sharpening} or resolution enhancement.  

The final non-standard step relates to the EMCD signal extraction itself.  Rather than applying a step-by-step approach, as is typically done in the EMCD literature, we instead opted for a more holistic procedure.  To this end, we have written an optimization routine in the Matlab programming language.  This routine optimizes several parameters in the extraction steps, by modelling the resulting EMCD signal as two pseudo-Voigt peaks and applying a least-squares approach. It is described at length in the methods section.  We believe that this results in a much more reliable and less bias-prone EMCD extraction and quantification.  The code for this approach is available along with the data used for this study on Zenodo \cite{thersleff_single-pass_2019} and we encourage the interested reader to explore this code on their own data and draw their own conclusions related to its reliability.

%Provided the reader accepts the data processing steps that are outlined in this paper, we can then move to the physical interpretation of the observed results.  Perhaps the most unexpected is the observation of the sign inversion on Fe $L_3$ for medium values of $q_y$.  To explain this, we turn to JAN ;-).  

%In addition to the theory, we emphasize here that the $q_y$ dimension in the $q-E$ diagram necessarily is a gross simplification.  It represents a 2D collapse of the $q_x$-projected CBED pattern projected onto the EELS spectrometer.  In other words, the regions referred to as "high $q_y$" in this paper actually represent an integration over only a small $q_x$ range.  In contrast, the lower $q_y$ values actually represent an integration over the entire $q_x$ range captured in the specific pass-through region of the aperture.  Moreover, all of these are convolved by the convergence of the probe, limiting the momentum resolution of the experiment even further to approximately the convergence semi-angle.  Thus, if - as predicted by theory - the real magnetic scattering contribution has both positive and negative contributions in reciprocal space, and if these are radially distributed, then all of these contributions will be superposed in the "low $q_y$" region of the $q-E$ diagram, whereas this superposition will not take place at higher $q_y$ values.  Therefore, we can expect that the "high $q_y$" region of the $q-E$ diagram better segregates the $q_x$ dimension and is thus less susceptible to a sign inversion.

In conclusion, we have presented our progress towards the use of a patterned aperture designed to allow for a single-pass STEM-EMCD experiment on the zone-axis of a magnetic crystal.  The analysis necessary to process these data spurred the development of considerable signal processing tools, which are published alongside this work \cite{thersleff_single-pass_2019}. %for development and need to be debated in the wider electron microscopy community.  
The use of these tools following the procedures justified here lead us to observe candidate EMCD signals that appear to be correlated to the alignment of the specimen along its zone axis as well as having a dependence on the non-dispersion dimension of the $q-E$ diagram, which we interpret in terms of theory and the geometry of the aperture / spectrometer coupling.  Thus, we adopt an optimistic outlook for this experimental design.  

%1) Discuss limitations: heavy data processing, model dependence, weak signal.  Also spectrometer artifacts and their limitations.  Discuss potential for post-processing removal of artifacts.
%
%2) Discuss advantages: zone-axis settings, small probe size (1-2nm?), single-pass, correlation of local grain orientation with spectroscopy (robustness with respect to orientation?)
%
%3) Discuss outlook: correlation of spatial coordinates with magnetic properties (grain boundaries, surfaces), probe can be reduced on a better machine (atomic resolution?)

\section*{Methods}

\subsection*{Sample fabrication}
The sample used for this experiment was prepared in the same manner as those used in Rusz et al. \cite{rusz_magnetic_2016} and Muto et al.\cite{muto_quantitative_2014}  A 10-nm-thick bcc Fe layer and a 3-nm-thick Al cap layer (to avoid oxidation of Fe) were deposited on 5-nm-thick Si$_3$N$_4$ membranes by thermal evaporation in an ultra-high vacuum molecular beam epitaxy (UHV-MBE) system. Thicknesses were controlled by calibrated quartz microbalances. We estimate relative thickness fluctuations of about 3\%, which ensures that no significant spectral intensity variation due to film thickness variation is expected. No ex situ or in situ preparation/cleaning was applied to the Si$_3$N$_4$ membranes prior to the deposition. The membranes were kept at room temperature during the deposition. The Fe films were post-annealed at 750$^{\circ}$C for 120 min. to increase the lateral Fe grain size to about 50~nm. The deposition of the Al cap layers was carried out at room temperature.  The disordered structure of the membranes (nanocrystalline or amorphous) led to a polycrystalline morphology of the metallic Fe/Al films. Air exposure after the deposition oxidised the Al cap layer to a depth of 1.5 -- 2~nm. Since the Al layer is 3~nm thick, a closed AlO$_x$ layer is maintained even in the presence of surface roughness (likely for a polycrystalline film). Some metallic Al may remain at the interface to the Fe film. As oxidation of the Fe film could have substantial effects on the intensity ratio of the $L_3$ and $L_2$ edges, the film was examined with EELS to probe if any oxidisation of the Fe took place before or after the EMCD measurements; using the fact that the oxygen K edge can be easily distinguished between aluminium and iron oxides. Nevertheless we found no iron oxides within the detection limit of EELS (< 1 at\%).  The oxygen edges visible in Fig.~\ref{fig:ApertureDispersion}b arise from averaging $q-E$ over all 2D EELS frames, including those from the background region (where no iron grain was present).

%It was grown by MBE on SiN and was post-annealed.  This yields a series of crystalline iron islands that are independently dispersed throughout the SiN substrate.  The thickness of these islands is around 30 nm.  They are capped with a 2~nm layer of Al, which converts to Al$_2$O$_3$ upon exposure to atmosphere to protect the iron from oxidation.

\subsection*{Experimental equipment}
The EMCD experiments were performed on a JEOL-2100F microscope operated at 200~kV.  The TEM is equipped with a Gatan Image Filter (GIF) Tridiem model using an UltraScan 1000 CCD camera, which was used to acquire the 2D EELS data.  Two additional CCD cameras are equipped at different column heights.  An Orius camera is fitted above the viewing chamber at the height of the JEOL HAADF detectors.  This camera runs at approximately 20 frames per second and has a large dynamic range, making it optimal for acquiring CBED patterns.  This camera was used for acquisition of the 4D STEM-Diffraction datacube as well as inspecting the CBED patterns without the aperture shadow.  The second CCD camera is an UltraScan 1000 situated between the viewing chamber and the GIF entrance.  This was used for positioning the aperture, as it has a much wider field of view than the UltraScan camera attached to the spectrometer.  

Image contrast in STEM mode was generated with two annular detectors and a secondary electron detector.  The first annular detector is located above the viewing chamber and was seldom used because of the strong demagnification of the diffraction patterns necessary for the EELS experiments.  The second annular detector is located at the spectrometer entrance.  This detector could be used due to the HAADF pass-through that was built into the aperture (see Fig.~\ref{fig:ApertureDesign}b).  However, since different camera lengths were needed for the 4D EELS and the STEM-diffraction experiments, yet the same area needed to be scanned, neither of these detectors was used for spatial registration.  Instead, a secondary electron detector mounted above the sample was turned on.  This position makes the contrast invariant to changes in the projector system of the TEM, ensuring high precision in the spatial registration of the probe between subsequent scans with different camera lengths. While the contrast was comparatively weak, the high probe current made it feasible to use for these experiments.

The TEM has a chamber directly below the viewing screen that is intended for film negatives.  This was emptied and the custom aperture system described above was inserted.  The aperture was positioned by the manual adjustment screws on the X-Y table (see Fig.~\ref{fig:ApertureDesign}) and its position with regards to the dispersion axis of the spectrometer was verified by inspecting its shadow on the CCD camera under parallel beam illumination.  It was also verified by looking at the unfocused zero-loss peak when operating the spectrometer in spectroscopy mode.  The aperture system was grounded to minimize charging and contamination effects, which were not observed even at high probe currents.

\subsection*{Data acquisition}

\subsubsection*{STEM alignment}

The condenser system in the TEM was aligned using custom settings tailored to this experiment.  The probe current was maximized by adjusting the ratio between the CL1 and CL3 lenses, while the convergence angle was set by adjusting the ratio between CL3 and the mini-condenser lens CM.  This is normally turned off in STEM mode for this instrument, but was manually engaged using the free lens control.  It was observed that higher convergence angles lead to a reduction of alignment quality, so an optimal balance was found at a convergence semiangle of 4.2~mrad and probe current of 2~nA.

Following base alignment, a survey image of the sample was taken at high camera length using the upper HAADF detector.  The probe was then moved by hand using the mouse in the Digiscan system and placed over candidate grains.  Simultaneously, the Orius camera was used to observe the CBED patterns.  The goal was to search for grains that were close to the Fe $\left [ 0 0 1 \right ]$ zone axis yet were large enough to demonstrate the scanning capabilities of this experiment.  Thickness was estimated by observing the extinction fringes in the CBED pattern.  The grain investigated in this paper required the goniometer to be slightly tilted to improve the orientation.

Diffraction conditions for the EELS experiments were configured using the projector system.  As precise rotation of the CBED pattern with respect to the dispersion plane of the spectrometer is critical for this experiment, and since this is a non-standard procedure, manual adjustment using the free lens control was required.  The rotation, focus, and demagnification of the CBED patterns were empirically optimized through manual adjustment of the four projector lenses in this TEM.  This was performed using the Orius camera with the probe stationary on a neighboring grain whose orientation respective to the region of interest was known and subsequently verified by briefly placing the probe on the region of interest (see Fig.~\ref{fig:ApertureDispersion}a).  Diffraction focus was maintained by scanning the probe over a large area and minimizing lateral translations (corresponding to beam tilt) with the projector lenses.

\subsubsection*{4D EELS acquisition}

Following optimization of the projector system, the spectrometer was switched to spectroscopy mode.  The shape of the zero-loss peak was optimized in 2D EELS mode using both focus and stigmator lenses in the spectrometer system.  Once the zero-loss peak shape was linear over a wide $q_y$ range, the drift tube was excited so that final tuning could be applied to the iron edges directly.

With the tuning completed, the sample was scanned.  A survey image was acquired using the secondary electron detector and a grid of $80 \times 80$ pixels was defined over the region of interest with a spacing of 6.3~nm.  A dwell time of 0.1~s was used and Digiscan was instructed to run in EDX mode yielding a scan speed of 8.7~pixels/s.  A custom script was then executed to record the time stamp at each pixel position using the Digital Micrograph Hook-up Script language \cite{schonstrom_time_2018}.  These time stamps were written to a Digital Micrograph persistent tag structure.  Simultaneously, the GIF CCD camera was switched to 2D EELS view mode and a second script was run that copied each frame to an empty data container \cite{hou_timed_2009}.  A time stamp for each frame copy was also recorded to a persistent tag structure, allowing for the spatial position of the 2D EELS image to be associated with the pixel position in Digiscan.  The exposure time on the camera was 0.1~s/frame which, including readout overhead, resulted in a data acquisition rate of 9.1 frames per second.  A slight undersampling of the 2D EELS frames with respect to the probe position occurred, and this was mollified by filling missing frames using nearest neighbor interpolation.  A spatial distribution of 2D EELS frames captured per pixel is provided in the supplementary information.

The spectrometer was set up for a dispersion of 0.2~eV and a vertical binning of 16.  This yields a 2D EELS image for each readout cycle of $32 \times 2048$ pixels, spanning an energy range of 420 -- 850~eV.  The camera length was adjusted as above to yield a $q_y$ range from -27 to 27~mrad spanning the vertical diameter of the aperture.  During acquisition, in a background thread, a sawtooth waveform was applied to the drift tube to continuously offset it \cite{hou_drift_2008,mitchell_drift_2017}.  The drift tube shifted the spectrum on the CCD camera by about 0.1 eV every second pixel.  The purpose of this was to perform binned \cite{bosman_optimizing_2008} and iterative \cite{schattschneider_iterative_1993} gain averaging.

\subsubsection*{STEM Diffraction acquisition}

After the 4D EELS experiment was recorded, the projector system was changed to yield an optimal camera length for the Orius CCD camera.  Critically, this very significant change to the projector settings has no influence on the contrast of the survey image generated with secondary electrons.  Hence, the same survey image could be used to correct for the probe starting position.  The same scripts used for recording time stamps in Digiscan \cite{schonstrom_time_2018} as well as recording the camera in view mode \cite{hou_timed_2009} were used to acquire all of the CBED patterns as the probe scanned across the region of interest.  In this case, the exposure time was lowered to 0.002~s/frame using binning 4 and the center quarter of the CCD camera.  This yielded a data acquisition rate of approximately 16 frames/s.

\subsection*{Data processing}

\subsubsection*{4D datacube construction}

Both the diffraction and 2D EELS data were stored as 3D image stacks in the Digital Micrograph format.  The recorded time stamps were used to assemble these data into 4D datacubes.  The timestamp at the beginning and end of each row was used to determine the range of images that would be assigned to that row.  These were then resampled to 80 images using a time vector generated with the timestamps and nearest neighbor interpolation.  The resulting spatial registration is visible in Fig.~\ref{fig:4Dscans}.  It should be noted that this method does not lead to perfect spatial registration.  Some scan jitter within each datacube is visible and the registration between datacubes is only accurate down to approximately 10 nm.  This source of error negatively impacts the mapping capabilities of this experiment, but should have minimal influence on the overall observation of EMCD when summing over larger areas.

\subsubsection*{4D EELS pretreatment}

Among the most important results of this study are the advances made in processing the 4D EELS datacubes.  All of the analysis was performed in custom-written Matlab code that is provided as supplementary information.  

First, pixel outliers (such as intense x-ray spikes) were removed by subtracting the moving average and then looking for deviations with a sigma of 5 or more.  Any spectra above this threshold were replaced by the mean of the four nearest neighbors.

After removing strong outliers, the 2D EELS spectra were shifted to align along the energy axis.  This procedure requires two steps.  In the first step, the energy offset due to the drift tube at each probe position was estimated.  The 2D EELS image from each pixel was vertically summed (summation along $q_y$), yielding an EELS spectrum image datacube $(x,y,\Delta E)$ datacube with dimensions $N_x \times N_y \times N_E$.  A reference spectrum was chosen and cross-correlated with all of the summed spectra over the Fe $L_{2,3}$ edges.  Outliers were detected by looking for large deviations from the sawtooth drift tube function and linearly interpolated over. The resulting energy shift map was then saved and applied to a copy of the raw 4D data on a frame-by-frame basis.  

This first step does not correct for shifts of the ionization edges along the $q_y$ dimension.  To correct for these, an adaptation of the method employed by Witjes et.~al for Raman spectra peak shift alignment is employed \cite{witjes_automatic_2000}.  The roughly aligned 4D EELS datacube (having dimensions $N_E \times N_{q_y} \times N_x \times N_y$) is rearranged into a 2D matrix having dimensions $N_E \times (N_{q_y} N_x N_y)$.  The implicit assumption here is that the entire dataset can be described as a series of 1D EELS spectra, which we denote as $S_k(E)$ where $k$ represents the linear index of each individual 1D spectrum.  We can assume that, in the absence of chemical shifts, the ionization edge of interest (Fe $L_3$ in this case) for all spectra should be exactly centered with the mean spectrum.  Since this is not the case, we can more accurately describe the EELS datacube as:

\begin{equation}
    S_k(E_j) = A_kf(E_j + \Delta E_k)
\end{equation}
where $A_k$ represents an amplitude for the $k^{\textup{th}}$ spectrum, $\Delta E_k$ represents the energy shift of the $k^{\textup{th}}$ spectrum and $E_j$ represents the energy vector of the mean.  Thus, $f(E_j)$ represents the mean spectrum that would be expected if all spectra were perfectly aligned.  We can now expand $S_k(E_j + \Delta E_k)$ in a Taylor series about $f(E_j)$:
\begin{equation}
    S_k(E_j) = A_k\left \{ f(E_j) + \Delta E_k \frac{\partial f}{\partial E} \Bigr|_{\substack{E_j}} + \frac{\Delta E_{k}^{2}}{2} \frac{\partial^2f}{\partial E^2} \Bigr|_{\substack{E_j}} + \cdots \right \}
\end{equation}
which can be rewritten as
\begin{equation}
    S_k(E_j) = b_{k,1}f(E_j) + b_{k,2}\frac{\partial f}{\partial E}\Bigr|_{\substack{E_j}} + b_{k,3}\frac{\partial^2 f}{\partial E^2} \Bigr|_{\substack{E_j}} + \cdots .
\end{equation}
In this case, $b_{k,i}$ denotes the $i^{\textup{th}}$ Taylor coefficient of the $k^{\textup{th}}$ spectrum.  Critically, the ratio between the second and first Taylor coefficients
\begin{equation}
    \frac{b_{k,2}}{b_{k,1}} = \frac{A_k \Delta E_k}{A_k} = \Delta E_k
\end{equation}
meaning that these coefficients can be used to estimate the energy shift for each individual spectrum.  The Taylor coefficients can thus be estimated simply by first estimating $f(E_j)$ to be the mean of $S_k(E_j)$ and writing this to the first column of a matrix $X$.  Subsequently, this spectrum is smoothed, numerically differentiated, and placed in the second column.  Using classical least squares regression
\begin{equation}
    B = (X^t X)^{-1} X^t S
\end{equation}
the Taylor coefficients and, thus, $\Delta E_k$ can be estimated.  This produces an energy shift correction that can be applied to each individual spectrum using Fourier shift theory to allow for sub-channel interpolation \cite{guizar-sicairos_efficient_2008}.  The procedure is iterated until $\Delta E_k$ becomes negligibly small.  It should be noted that the higher order Taylor coefficients can be used to correct for additional aberrations, such as peak broadening \cite{witjes_automatic_2000-1,witjes_modelling_2001}.  However, this becomes increasingly difficult in the limit of strong noise corruption, such as is present in this dataset.

The value of the energy shift $\Delta E_k$ for each of these steps was saved and then added together.  The final shift correction was then applied to the raw data, resulting in the optimal energy alignment.  Subsequently, the 4D EELS datacube was truncated within the energy range 650 -- 830 eV.  This marks the end of the pretreatment stage for spectral data.

\subsubsection*{4D STEM DP pretreatment}

The data processing workflow for the 4D STEM DP datacube is much less involved than for the 4D EELS datacube.  Following interpolation into a 4D datacube, the data were visually explored using the PyXem software package \cite{johnstone_pyxem/pyxem:_2019}, which is an extention to Hyperspy \cite{francisco_de_la_pena_hyperspy/hyperspy:_2018}.  This tool was used to produce qualitative orientation maps using virtual dark field images generated by placing a virtual aperture over the (0~1~1) Kikuchi line bands in the outer regions of the diffraction patterns.

%An animated visualization of the full dataset is provided in the supplementary information in the form of a movie.

\subsubsection*{EMCD signal extraction}

The extraction of EMCD signals was performed using the \texttt{fmincon} function implemented in the optimization toolbox in the Matlab programming language.  An objective function was written that contains the following steps.  

First, the desired $q$-range and orientation range was determined and single EELS spectra for both chiral plus and chiral minus were obtained by integrating over these regions of interest.  

Second, the pre-edge background was modelled in the energy range 650--700 eV using an inverse power law
\begin{equation}
    \label{eq:PreEdgeBG}
    f_{\pm,\textup{BG}}(E) = f_{\pm}(E) - A_{\pm}E^{r_{\pm}}
\end{equation}
where $f_{\pm}(E)$ represents the raw extracted signal for chiral plus or minus, $f_{\pm,BG}(E)$ is the background-subtracted signal, $E$ is the energy vector (spanning the range 650--700 eV), and the remaining parameters are defined in Table~\ref{tab:fminconConstraints}.  This model was subtracted from the raw spectra as shown in Eq.~\ref{eq:PreEdgeBG}.

In many cases, we noticed that the peak width of the chiral plus and chiral minus spectra was different.  This yields a complex up-down-up EELS difference signal reminiscent of what one would expect for magnetite on the Fe $L_3$ edge.  Inspection of the spectra indicates that this was almost certainly caused by residual spectral aberrations resulting from imperfect tuning of the $S_x$ and $S_y$ lenses in the spectrometer.  Correction of these aberrations to the precision demanded by this EMCD experiment is likely not possible without computer assistance.  To counteract this effect, we employ a profile matching routine at this stage of the objective function.  The routine 
\begin{equation}
    f_{-,s}(E) = f_{-,\textup{BG}}(E) - k_s\frac{\partial^2 f_{-,BG}(E)}{\partial E^2}
\end{equation}
takes the second derivative of the smoothed chiral minus signal, scales it by $k_s$, and subtracts it from the original signal, yielding the same spectrum with a different peak width $f_{-,s}(E)$.  $k_s$ is treated as a parameter that is allowed to vary during optimization.  Negative values of this scalar will result in a peak broadening, emphasizing that the intent is to ensure that the peak shapes \emph{match} rather than to modify the experimental energy resolution via peak sharpening.  We also observe that the cumulative sum of the second derivative shows a zero crossing around 718~eV, indicating that the area underneath the modified peak is the same as before this procedure.  A figure demonstrating this is presented in the supplementary information.  Once the peak shapes are made similar to each other, a very small mismatch in the alignment of the chiral plus and chiral minus spectra was occasionally observed, usually on the order of 0.1 channels.  We thus also shift the chiral minus by a non-integer amount $\Delta E$ using Fourier shift theory.  The shift amount is also parameterized in the objective function and allowed to vary when performing the optimization.

For the fourth step, a post-edge normalization was performed.  The parameters for the post-edge normalization were determined by first computing the energy-dependent ratio between the background subtracted spectra, $D(E)$
\begin{equation}
    D(E) = \frac{f_{+,\textup{BG}}(E)}{f_{-,s}(E)} .
\end{equation}
A linear regression line with slope $m$ and intercept $d$ was fit to the post-edge ratios in the energy range 730--760 eV.  This line was extrapolated over the entire background subtracted spectra and then multiplied by the chiral minus signal to normalize it to chiral plus.  %As noted in the discussion section, this approach has been published previously \cite{schneider_magnetic_2016}.  Matlab code for this approach is provided in the supplementary information.

Following the post-edge normalization, the spectra were normalized to the maximum value of either the chiral plus or chiral minus spectra and subtracted from each other, yielding $f_\textup{EMCD}(E)$.  The normalization here is largely done for aesthetic purposes and simplifies the interpretation of the EMCD signal fitting parameters.  This step has no impact on quantitative values extracted from EMCD spectra.

The sixth step entails the introduction of a model for the EMCD signal itself.  Since this experiment was performed on bcc iron, we use a simple model of two pseudo-Voigt peaks 
\begin{equation}
\begin{split}
    f_{\textup{fit}}(E) = \left [ a_{1}\eta \left\{ 1 + \left ( \frac{E - c_1}{b_{1a} + b_{1b}(E-c_1)} \right)^2 \right\}^{-1}  + a_{1}(1-\eta)  \exp \left\{ -\ln(2) \left ( \frac{E-c_1}{b_{1a} + b_{1b}(E-c_1)} \right)^2  \right\} \right]_{L_3} + \\
    \left [ a_{2} \eta \left\{ 1 + \left ( \frac{E - c_2}{b_{2a} + b_{2b}(E-c_2)} \right)^2 \right\}^{-1}  + a_{2}(1-\eta)  \exp \left\{  -\ln(2) \left( \frac{E-c_2}{b_{2a} + b_{2b}(E-c_2)} \right)^2  \right\} \right ]_{L_2}
\end{split}
\end{equation}
where all the parameters are defined in Table~\ref{tab:fminconConstraints}.  Note that $a_1$ and $a_2$ are constrained to have opposite sign.  This model adds nine parameters to the optimization routine.  The model is subtracted from $f_\textup{EMCD}(E)$, yielding residual errors.  The objective function passed to \texttt{fmincon} minimizes the sum of the square of these errors.

The above steps lead to a total of 17 parameters that are allowed to vary by the \texttt{fmincon} function.  These parameters were constrained to ranges listed in table \ref{tab:fminconConstraints}.  The errors arising from the pre-edge background range for both chiral plus and chiral minus were passed separately, so as to allow the optimization to favor a good fit to the input spectra rather than the subtracted result.  The code used for this approach can be downloaded from Zenodo \cite{thersleff_single-pass_2019} and the full analysis and figure generation script for this manuscript is formatted for publication in the supplementary information.

\begin{table}[ht]
	\centering
	\begin{tabular}{|l|l|l|l|}
		\hline
		\textbf{Parameter Description} & \textbf{Symbol} & \textbf{Lower bound} & \textbf{Upper bound} \\
		\hline
		Chiral plus pre-edge amplitude & $A_+$ & 0 & +$\infty$ \\
		\hline
		Chiral minus pre-edge amplitude & $A_-$ & 0 & +$\infty$ \\
		\hline
		Chiral plus pre-edge slope & $r_+$ & -5 & 0 \\
		\hline
		Chiral minus pre-edge slope & $r_-$ & -5 & 0 \\
		\hline
		Chiral minus sharpening scalar & $k_s$ & -4 & 4 \\
		\hline
		Chiral minus shift & $\Delta E$ & -3 & 3 \\
		\hline
		Post-edge normalization slope & $m$ & -$\infty$ & +$\infty$ \\
		\hline
		Post-edge normalization intercept & $d$ & -$\infty$ & +$\infty$ \\
		\hline
		EMCD amplitude $L_3$ & $a_1$ & 0 & +$\infty$ \\
		\hline
		EMCD broadening a $L_3$ [eV] & $b_{1a}$ & 0 & 5.0 \\
		\hline
		EMCD broadening b $L_3$ [eV] & $b_{1b}$ & 0 & 5.0 \\
		\hline
		EMCD center $L_3$ [eV] & $c_1$ & 700 & 712 \\
		\hline
		EMCD amplitude $L_2$ & $a_2$ & -$\infty$ & 0 \\
		\hline
		EMCD broadening a $L_2$ [eV] & $b_{2a}$ & 0 & 5.0 \\
		\hline
		EMCD broadening b $L_2$ [eV] & $b_{2b}$ & 0 & 5.0 \\
		\hline
		EMCD center $L_2$ [eV] & $c_2$ & 716 & 725 \\
		\hline
		Lorentzian / Gaussian mixing parameter & $\eta$ & 0 & 1 \\
		\hline
	\end{tabular}
	\caption{Constraints for the parameters passed to \texttt{fmincon} in the EMCD signal extraction.  Note that the EMCD amplitudes were not constrained to be positive and negative as above; rather, they were constrained to have opposite sign from each other.}
	\label{tab:fminconConstraints}
\end{table}

%\subsection*{Simulations(?)}
%
%MATS.v2, parameters, etc.

\section*{Data Availability}

The data in the manuscript are available for download under the Creative Commons Attribution 4.0 International license \cite{thersleff_single-pass_2019}.  The code used for generating the figures including the EMCD optimization routines is likewise available under the GNU public license 3.0.  This code is presented in a human-readable format as the supplementary information to this manuscript.

\bibliography{PAPER_Patterned-Aperture}

\begin{thebibliography}{10}
\urlstyle{rm}
\expandafter\ifx\csname url\endcsname\relax
  \def\url#1{\texttt{#1}}\fi
\expandafter\ifx\csname urlprefix\endcsname\relax\def\urlprefix{URL }\fi
\expandafter\ifx\csname doiprefix\endcsname\relax\def\doiprefix{DOI: }\fi
\providecommand{\bibinfo}[2]{#2}
\providecommand{\eprint}[2][]{\url{#2}}

\bibitem{chao_soft_2005}
\bibinfo{author}{Chao, W.}, \bibinfo{author}{Harteneck, B.~D.},
  \bibinfo{author}{Liddle, J.~A.}, \bibinfo{author}{Anderson, E.~H.} \&
  \bibinfo{author}{Attwood, D.~T.}
\newblock \bibinfo{journal}{\bibinfo{title}{Soft {X}-ray microscopy at a
  spatial resolution better than 15 nm}}.
\newblock {\emph{\JournalTitle{Nature}}} \textbf{\bibinfo{volume}{435}},
  \bibinfo{pages}{1210--1213},
  \doiprefix\url{https://doi.org/10.1038/nature03719} (\bibinfo{year}{2005}).

\bibitem{wiesendanger_topographic_1992}
\bibinfo{author}{Wiesendanger, R.} \emph{et~al.}
\newblock \bibinfo{journal}{\bibinfo{title}{Topographic and
  {Magnetic}-{Sensitive} {Scanning} {Tunneling} {Microscope} {Study} of
  {Magnetite}}}.
\newblock {\emph{\JournalTitle{Science}}} \textbf{\bibinfo{volume}{255}},
  \bibinfo{pages}{583--586},
  \doiprefix\url{https://doi.org/10.1126/science.255.5044.583}
  (\bibinfo{year}{1992}).

\bibitem{heinze_real-space_2000}
\bibinfo{author}{Heinze, S.} \emph{et~al.}
\newblock \bibinfo{journal}{\bibinfo{title}{Real-{Space} {Imaging} of
  {Two}-{Dimensional} {Antiferromagnetism} on the {Atomic} {Scale}}}.
\newblock {\emph{\JournalTitle{Science}}} \textbf{\bibinfo{volume}{288}},
  \bibinfo{pages}{1805--1808},
  \doiprefix\url{https://doi.org/10.1126/science.288.5472.1805}
  (\bibinfo{year}{2000}).

\bibitem{kaiser_magnetic_2007}
\bibinfo{author}{Kaiser, U.}, \bibinfo{author}{Schwarz, A.} \&
  \bibinfo{author}{Wiesendanger, R.}
\newblock \bibinfo{journal}{\bibinfo{title}{Magnetic exchange force microscopy
  with atomic resolution}}.
\newblock {\emph{\JournalTitle{Nature}}} \textbf{\bibinfo{volume}{446}},
  \bibinfo{pages}{522--525},
  \doiprefix\url{https://doi.org/10.1038/nature05617} (\bibinfo{year}{2007}).

\bibitem{midgley_electron_2009}
\bibinfo{author}{Midgley, P.~A.} \& \bibinfo{author}{Dunin-Borkowski, R.~E.}
\newblock \bibinfo{journal}{\bibinfo{title}{Electron tomography and holography
  in materials science}}.
\newblock {\emph{\JournalTitle{Nature Materials}}}
  \textbf{\bibinfo{volume}{8}}, \bibinfo{pages}{271--280},
  \doiprefix\url{https://doi.org/10.1038/nmat2406} (\bibinfo{year}{2009}).

\bibitem{schattschneider_detection_2006}
\bibinfo{author}{Schattschneider, P.} \emph{et~al.}
\newblock \bibinfo{journal}{\bibinfo{title}{Detection of magnetic circular
  dichroism using a transmission electron microscope}}.
\newblock {\emph{\JournalTitle{Nature}}} \textbf{\bibinfo{volume}{441}},
  \bibinfo{pages}{486--488},
  \doiprefix\url{https://doi.org/10.1038/nature04778} (\bibinfo{year}{2006}).

\bibitem{hebert_proposal_2003}
\bibinfo{author}{Hébert, C.} \& \bibinfo{author}{Schattschneider, P.}
\newblock \bibinfo{journal}{\bibinfo{title}{A proposal for dichroic experiments
  in the electron microscope}}.
\newblock {\emph{\JournalTitle{Ultramicroscopy}}}
  \textbf{\bibinfo{volume}{96}}, \bibinfo{pages}{463--468},
  \doiprefix\url{https://doi.org/10.1016/S0304-3991(03)00108-6}
  (\bibinfo{year}{2003}).

\bibitem{schattschneider_detection_2008}
\bibinfo{author}{Schattschneider, P.} \emph{et~al.}
\newblock \bibinfo{journal}{\bibinfo{title}{Detection of magnetic circular
  dichroism on the two-nanometer scale}}.
\newblock {\emph{\JournalTitle{Phys. Rev. B}}} \textbf{\bibinfo{volume}{78}},
  \doiprefix\url{https://doi.org/10.1103/PhysRevB.78.104413}
  (\bibinfo{year}{2008}).

\bibitem{salafranca_surfactant_2012}
\bibinfo{author}{Salafranca, J.} \emph{et~al.}
\newblock \bibinfo{journal}{\bibinfo{title}{Surfactant {Organic} {Molecules}
  {Restore} {Magnetism} in {Metal}-{Oxide} {Nanoparticle} {Surfaces}}}.
\newblock {\emph{\JournalTitle{Nano Lett.}}} \textbf{\bibinfo{volume}{12}},
  \bibinfo{pages}{2499--2503},
  \doiprefix\url{https://doi.org/10.1021/nl300665z} (\bibinfo{year}{2012}).

\bibitem{thersleff_quantitative_2015}
\bibinfo{author}{Thersleff, T.} \emph{et~al.}
\newblock \bibinfo{journal}{\bibinfo{title}{Quantitative analysis of magnetic
  spin and orbital moments from an oxidized iron (1 1 0) surface using electron
  magnetic circular dichroism}}.
\newblock {\emph{\JournalTitle{Sci. Rep.}}} \textbf{\bibinfo{volume}{5}},
  \bibinfo{pages}{13012}, \doiprefix\url{https://doi.org/10.1038/srep13012}
  (\bibinfo{year}{2015}).

\bibitem{schattschneider_mapping_2012}
\bibinfo{author}{Schattschneider, P.}, \bibinfo{author}{Schaffer, B.},
  \bibinfo{author}{Ennen, I.} \& \bibinfo{author}{Verbeeck, J.}
\newblock \bibinfo{journal}{\bibinfo{title}{Mapping spin-polarized transitions
  with atomic resolution}}.
\newblock {\emph{\JournalTitle{Phys. Rev. B}}} \textbf{\bibinfo{volume}{85}},
  \bibinfo{pages}{134422},
  \doiprefix\url{https://doi.org/10.1103/PhysRevB.85.134422}
  (\bibinfo{year}{2012}).

\bibitem{thersleff_towards_2017}
\bibinfo{author}{Thersleff, T.} \emph{et~al.}
\newblock \bibinfo{journal}{\bibinfo{title}{Towards sub-nanometer real-space
  observation of spin and orbital magnetism at the {Fe}/{MgO} interface}}.
\newblock {\emph{\JournalTitle{Sci. Rep.}}} \textbf{\bibinfo{volume}{7}},
  \bibinfo{pages}{44802}, \doiprefix\url{https://doi.org/10.1038/srep44802}
  (\bibinfo{year}{2017}).

\bibitem{thersleff_detection_2016}
\bibinfo{author}{Thersleff, T.}, \bibinfo{author}{Rusz, J.},
  \bibinfo{author}{Hjörvarsson, B.} \& \bibinfo{author}{Leifer, K.}
\newblock \bibinfo{journal}{\bibinfo{title}{Detection of magnetic circular
  dichroism with subnanometer convergent electron beams}}.
\newblock {\emph{\JournalTitle{Phys. Rev. B}}} \textbf{\bibinfo{volume}{94}},
  \bibinfo{pages}{134430},
  \doiprefix\url{https://doi.org/10.1103/PhysRevB.94.134430}
  (\bibinfo{year}{2016}).

\bibitem{rusz_magnetic_2016}
\bibinfo{author}{Rusz, J.} \emph{et~al.}
\newblock \bibinfo{journal}{\bibinfo{title}{Magnetic measurements with
  atomic-plane resolution}}.
\newblock {\emph{\JournalTitle{Nature Communications}}}
  \textbf{\bibinfo{volume}{7}}, \bibinfo{pages}{12672},
  \doiprefix\url{https://doi.org/10.1038/ncomms12672} (\bibinfo{year}{2016}).

\bibitem{song_detection_2016}
\bibinfo{author}{Song, D.}, \bibinfo{author}{Rusz, J.}, \bibinfo{author}{Cai,
  J.} \& \bibinfo{author}{Zhu, J.}
\newblock \bibinfo{journal}{\bibinfo{title}{Detection of electron magnetic
  circular dichroism signals under zone axial diffraction geometry}}.
\newblock {\emph{\JournalTitle{Ultramicroscopy}}}
  \textbf{\bibinfo{volume}{169}}, \bibinfo{pages}{44--54},
  \doiprefix\url{https://doi.org/10.1016/j.ultramic.2016.07.005}
  (\bibinfo{year}{2016}).

\bibitem{spiegelberg_blind_2018}
\bibinfo{author}{Spiegelberg, J.}, \bibinfo{author}{Song, D.},
  \bibinfo{author}{Dunin-Borkowski, R.~E.}, \bibinfo{author}{Zhu, J.} \&
  \bibinfo{author}{Rusz, J.}
\newblock \bibinfo{journal}{\bibinfo{title}{Blind identification of magnetic
  signals in electron magnetic chiral dichroism using independent component
  analysis}}.
\newblock {\emph{\JournalTitle{Ultramicroscopy}}}
  \textbf{\bibinfo{volume}{195}}, \bibinfo{pages}{129--135},
  \doiprefix\url{https://doi.org/10.1016/j.ultramic.2018.08.021}
  (\bibinfo{year}{2018}).

\bibitem{rusz_achieving_2014}
\bibinfo{author}{Rusz, J.}, \bibinfo{author}{Idrobo, J.-C.} \&
  \bibinfo{author}{Bhowmick, S.}
\newblock \bibinfo{journal}{\bibinfo{title}{Achieving {Atomic} {Resolution}
  {Magnetic} {Dichroism} by {Controlling} the {Phase} {Symmetry} of an
  {Electron} {Probe}}}.
\newblock {\emph{\JournalTitle{Phys. Rev. Lett.}}}
  \textbf{\bibinfo{volume}{113}}, \bibinfo{pages}{145501},
  \doiprefix\url{https://doi.org/10.1103/PhysRevLett.113.145501}
  (\bibinfo{year}{2014}).

\bibitem{idrobo_detecting_2016}
\bibinfo{author}{Idrobo, J.~C.} \emph{et~al.}
\newblock \bibinfo{journal}{\bibinfo{title}{Detecting magnetic ordering with
  atomic size electron probes}}.
\newblock {\emph{\JournalTitle{Advanced Structural and Chemical Imaging}}}
  \textbf{\bibinfo{volume}{2}}, \bibinfo{pages}{5},
  \doiprefix\url{https://doi.org/10.1186/s40679-016-0019-9}
  (\bibinfo{year}{2016}).

\bibitem{wang_atomic_2018}
\bibinfo{author}{Wang, Z.} \emph{et~al.}
\newblock \bibinfo{journal}{\bibinfo{title}{Atomic scale imaging of magnetic
  circular dichroism by achromatic electron microscopy}}.
\newblock {\emph{\JournalTitle{Nature Materials}}} \bibinfo{pages}{1},
  \doiprefix\url{https://doi.org/10.1038/s41563-017-0010-4}
  (\bibinfo{year}{2018}).

\bibitem{negi_proposal_2019}
\bibinfo{author}{Negi, D.} \emph{et~al.}
\newblock \bibinfo{journal}{\bibinfo{title}{Proposal for {Measuring}
  {Magnetism} with {Patterned} {Apertures} in a {Transmission} {Electron}
  {Microscope}}}.
\newblock {\emph{\JournalTitle{Phys. Rev. Lett.}}}
  \textbf{\bibinfo{volume}{122}}, \bibinfo{pages}{037201},
  \doiprefix\url{https://doi.org/10.1103/PhysRevLett.122.037201}
  (\bibinfo{year}{2019}).

\bibitem{schattschneider_magnetic_2008}
\bibinfo{author}{Schattschneider, P.} \emph{et~al.}
\newblock \bibinfo{journal}{\bibinfo{title}{Magnetic circular dichroism in
  {EELS}: {Towards} 10nm resolution}}.
\newblock {\emph{\JournalTitle{Ultramicroscopy}}}
  \textbf{\bibinfo{volume}{108}}, \bibinfo{pages}{433--438},
  \doiprefix\url{https://doi.org/10.1016/j.ultramic.2007.07.002}
  (\bibinfo{year}{2008}).

\bibitem{ali_quantitative_2018}
\bibinfo{author}{Ali, H.}, \bibinfo{author}{Warnatz, T.}, \bibinfo{author}{Xie,
  L.}, \bibinfo{author}{Hjörvarsson, B.} \& \bibinfo{author}{Leifer, K.}
\newblock \bibinfo{journal}{\bibinfo{title}{Quantitative {EMCD} by use of a
  double aperture for simultaneous acquisition of {EELS}}}.
\newblock {\emph{\JournalTitle{Ultramicroscopy}}}
  \doiprefix\url{https://doi.org/10.1016/j.ultramic.2018.10.012}
  (\bibinfo{year}{2018}).

\bibitem{thersleff_single-pass_2019}
\bibinfo{author}{Thersleff, T.} \emph{et~al.}
\newblock \bibinfo{title}{Single-pass {STEM}-{EMCD} on a zone axis using
  apatterned aperture: {Data} and methodology},
  \doiprefix\url{https://doi.org/10.5281/zenodo.3361582}
  (\bibinfo{year}{2019}).

\bibitem{francisco_de_la_pena_hyperspy/hyperspy:_2018}
\bibinfo{author}{de~la Peña, F.} \emph{et~al.}
\newblock \bibinfo{title}{hyperspy/hyperspy: {HyperSpy} 1.3.1},
  \doiprefix\url{https://doi.org/10.5281/zenodo.1221347}
  (\bibinfo{year}{2018}).

\bibitem{johnstone_pyxem/pyxem:_2019}
\bibinfo{author}{Johnstone, D.~N.} \emph{et~al.}
\newblock \bibinfo{title}{pyxem/pyxem: {pyXem} 0.7.1},
  \doiprefix\url{https://doi.org/10.5281/zenodo.2649351}
  (\bibinfo{year}{2019}).

\bibitem{schonstrom_time_2018}
\bibinfo{author}{Schönström, L.}
\newblock \bibinfo{title}{Time {Stamp} {Code}} (\bibinfo{year}{2018}).
\newblock \bibinfo{note}{This code is available upon request}.

\bibitem{rusz_asymmetry_2010}
\bibinfo{author}{Rusz, J.}, \bibinfo{author}{Oppeneer, P.},
  \bibinfo{author}{Lidbaum, H.}, \bibinfo{author}{Rubino, S.} \&
  \bibinfo{author}{Leifer, K.}
\newblock \bibinfo{journal}{\bibinfo{title}{Asymmetry of the two-beam geometry
  in {EMCD} experiments}}.
\newblock {\emph{\JournalTitle{J. Microsc.}}} \textbf{\bibinfo{volume}{237}},
  \bibinfo{pages}{465--468},
  \doiprefix\url{https://doi.org/10.1111/j.1365-2818.2009.03295.x}
  (\bibinfo{year}{2010}).

\bibitem{song_effect_2014}
\bibinfo{author}{Song, D.}, \bibinfo{author}{Wang, Z.} \& \bibinfo{author}{Zhu,
  J.}
\newblock \bibinfo{journal}{\bibinfo{title}{Effect of the asymmetry of
  dynamical electron diffraction on intensity of acquired {EMCD} signals}}.
\newblock {\emph{\JournalTitle{Ultramicroscopy}}}
  \textbf{\bibinfo{volume}{148}}, \bibinfo{pages}{42--51},
  \doiprefix\url{https://doi.org/10.1016/j.ultramic.2014.08.012}
  (\bibinfo{year}{2014}).

\bibitem{schneider_magnetic_2016}
\bibinfo{author}{Schneider, S.} \emph{et~al.}
\newblock \bibinfo{journal}{\bibinfo{title}{Magnetic properties of single
  nanomagnets: {Electron} energy-loss magnetic chiral dichroism on {FePt}
  nanoparticles}}.
\newblock {\emph{\JournalTitle{Ultramicroscopy}}}
  \textbf{\bibinfo{volume}{171}}, \bibinfo{pages}{186--194},
  \doiprefix\url{https://doi.org/10.1016/j.ultramic.2016.09.009}
  (\bibinfo{year}{2016}).

\bibitem{muto_quantitative_2014}
\bibinfo{author}{Muto, S.} \emph{et~al.}
\newblock \bibinfo{journal}{\bibinfo{title}{Quantitative characterization of
  nanoscale polycrystalline magnets with electron magnetic circular
  dichroism}}.
\newblock {\emph{\JournalTitle{Nat. Commun.}}} \textbf{\bibinfo{volume}{5}},
  \bibinfo{pages}{3138}, \doiprefix\url{https://doi.org/10.1038/ncomms4138}
  (\bibinfo{year}{2014}).

\bibitem{hou_timed_2009}
\bibinfo{author}{Hou, V.} \& \bibinfo{author}{Schönström, L.}
\newblock \bibinfo{title}{Timed {Image} {Recorder}} (\bibinfo{year}{2009}).

\bibitem{hou_drift_2008}
\bibinfo{author}{Hou, V.}
\newblock \bibinfo{title}{Drift {Tube} {Scan}} (\bibinfo{year}{2008}).

\bibitem{mitchell_drift_2017}
\bibinfo{author}{Mitchell, D.~R.}
\newblock \bibinfo{title}{Drift {Tube} {Scanning}} (\bibinfo{year}{2017}).

\bibitem{bosman_optimizing_2008}
\bibinfo{author}{Bosman, M.} \& \bibinfo{author}{Keast, V.~J.}
\newblock \bibinfo{journal}{\bibinfo{title}{Optimizing {EELS} acquisition}}.
\newblock {\emph{\JournalTitle{Ultramicroscopy}}}
  \textbf{\bibinfo{volume}{108}}, \bibinfo{pages}{837--846},
  \doiprefix\url{https://doi.org/10.1016/j.ultramic.2008.02.003}
  (\bibinfo{year}{2008}).

\bibitem{schattschneider_iterative_1993}
\bibinfo{author}{Schattschneider, P.} \& \bibinfo{author}{Jonas, P.}
\newblock \bibinfo{journal}{\bibinfo{title}{Iterative reduction of gain
  variations in parallel electron energy loss spectrometry}}.
\newblock {\emph{\JournalTitle{Ultramicroscopy}}}
  \textbf{\bibinfo{volume}{49}}, \bibinfo{pages}{179--188},
  \doiprefix\url{https://doi.org/10.1016/0304-3991(93)90224-L}
  (\bibinfo{year}{1993}).

\bibitem{witjes_automatic_2000}
\bibinfo{author}{Witjes, H.}, \bibinfo{author}{van~den Brink, M.},
  \bibinfo{author}{Melssen, W.} \& \bibinfo{author}{Buydens, L.}
\newblock \bibinfo{journal}{\bibinfo{title}{Automatic correction of peak shifts
  in {Raman} spectra before {PLS} regression}}.
\newblock {\emph{\JournalTitle{Chemometrics and Intelligent Laboratory
  Systems}}} \textbf{\bibinfo{volume}{52}}, \bibinfo{pages}{105--116},
  \doiprefix\url{https://doi.org/10.1016/S0169-7439(00)00085-X}
  (\bibinfo{year}{2000}).

\bibitem{guizar-sicairos_efficient_2008}
\bibinfo{author}{Guizar-Sicairos, M.}, \bibinfo{author}{Thurman, S.~T.} \&
  \bibinfo{author}{Fienup, J.~R.}
\newblock \bibinfo{journal}{\bibinfo{title}{Efficient subpixel image
  registration algorithms}}.
\newblock {\emph{\JournalTitle{Opt. Lett., OL}}} \textbf{\bibinfo{volume}{33}},
  \bibinfo{pages}{156--158},
  \doiprefix\url{https://doi.org/10.1364/OL.33.000156} (\bibinfo{year}{2008}).

\bibitem{witjes_automatic_2000-1}
\bibinfo{author}{Witjes, H.} \emph{et~al.}
\newblock \bibinfo{journal}{\bibinfo{title}{Automatic {Correction} for {Phase}
  {Shifts}, {Frequency} {Shifts}, and {Lineshape} {Distortions} across a
  {Series} of {Single} {Resonance} {Lines} in {Large} {Spectral} {Data}
  {Sets}}}.
\newblock {\emph{\JournalTitle{Journal of Magnetic Resonance}}}
  \textbf{\bibinfo{volume}{144}}, \bibinfo{pages}{35--44},
  \doiprefix\url{https://doi.org/10.1006/jmre.2000.2021}
  (\bibinfo{year}{2000}).

\bibitem{witjes_modelling_2001}
\bibinfo{author}{Witjes, H.}, \bibinfo{author}{Pepers, M.},
  \bibinfo{author}{Melssen, W.~J.} \& \bibinfo{author}{Buydens, L. M.~C.}
\newblock \bibinfo{journal}{\bibinfo{title}{Modelling phase shifts, peak shifts
  and peak width variations in spectral data sets: its value in multivariate
  data analysis}}.
\newblock {\emph{\JournalTitle{Analytica Chimica Acta}}}
  \textbf{\bibinfo{volume}{432}}, \bibinfo{pages}{113--124},
  \doiprefix\url{https://doi.org/10.1016/S0003-2670(00)01349-0}
  (\bibinfo{year}{2001}).

\end{thebibliography}

%\noindent LaTeX formats citations and references automatically using the bibliography records in your .bib file, which you can edit via the project menu. Use the cite command for an inline citation, e.g.  \cite{Hao:gidmaps:2014}.

%For data citations of datasets uploaded to e.g. \emph{figshare}, please use the \verb|howpublished| option in the bib entry to specify the platform and the link, as in the \verb|Hao:gidmaps:2014| example in the sample bibliography file.

\section*{Acknowledgements}

T.~T.~acknowledges Wei Wan for suggesting to install the aperture device in the negative chamber of the TEM, Sebastian Schneider and Carolin Schmitz-Antoniak for the formative discussions on post-edge normalization routines, as well as Stef Smeets and Bin Wang for the fruitful scientific exchanges and inspiring discussions regarding open science publishing.  L.~S.~acknowledges Jakob Paulin for valuable contributions to the design and build of the aperture holding system.  T.~T.~and L.~S.~acknowledge funding from the Swedish Research Council (project nr. 2016-05113).  C.-W.~T.~and T.~T.~acknowledge funding from the Swedish Strategic Research foundation (project nr. ITM17-0301).  J.~R. acknowledges funding from the Swedish Research Council (project nr. 2017-04026).

\section*{Author contributions statement}

T.~T.~proposed the single-pass mirror symmetry geometry investigated in this manuscript, which was based off of the original idea for using a patterned aperture by J.~R.\\
T.~T.~and L.~S.~designed the experiment, the aperture plate and the aperture holding system together with valuable input from S.~M and C.~-W.~T.\\
L.~S.~authored the hook-up script allowing for 2D-EELS and STEM-DP frame capture, and built the aperture holding system.\\
T.~T.~acquired the data and performed the analysis, including authoring all of the relevant Matlab code (except where otherwise noted).\\
R.~A., D.~B., and C.~S.~grew and prepared the iron sample used in this manuscript.\\
All authors contributed to the writing of the manuscript and the interpretation of the results.

\section*{Additional information}

\subsection*{Competing Interests}
The authors declare no competing interests.

%To include, in this order: \textbf{Accession codes} (where applicable); \textbf{Competing interests} (mandatory statement). 

%The corresponding author is responsible for submitting a \href{http://www.nature.com/srep/policies/index.html#competing}{competing interests statement} on behalf of all authors of the paper. This statement must be included in the submitted article file.

%Figures and tables can be referenced in LaTeX using the ref command, e.g. Figure \ref{fig:stream} and Table \ref{tab:example}.

\end{document}